\documentclass{llncs}

\usepackage{makeidx}

\usepackage{rotating}
\usepackage{cite}
\usepackage[usenames]{color}
\usepackage{graphicx}
\usepackage{algorithmicx}
\usepackage{times}

\DeclareGraphicsExtensions{.pdf}
\DeclareGraphicsExtensions{.jpg}
\DeclareGraphicsExtensions{.eps}
\usepackage{amsmath,amssymb}

\usepackage{amsthm}
\usepackage{amssymb}
\usepackage{amscd}
\usepackage{color}
\usepackage{fancybox}
\usepackage{eepic}
\usepackage{graphicx}
\usepackage{xspace}
\usepackage[subrefformat=parens,labelformat=parens]{subfig}
\usepackage{hhline}
\usepackage{algpseudocode} 
\usepackage{url}
\usepackage{times}
\newcommand{\id}{{\it id}\xspace}

\newcommand{\Fig}[1]{Fig.~\ref{#1}\xspace}
\newcommand{\SubFig}[1]{Fig.~\subref*{#1}\xspace}
\newcommand{\Sec}[1]{Section~\ref{#1}\xspace}

\newcommand{\IndState}[1][1]{\State\hspace{#1\algorithmicindent}}
\newcommand{\Shared}{\ensuremath{Shared}\xspace}
\newcommand{\Invalid}{\ensuremath{Invalid}\xspace}
\newcommand{\Exclusive}{\ensuremath{Exclusive}\xspace}
\newcommand{\Invalidate}{\ensuremath{Invalidate}\xspace}
\newcommand{\curptr}{\ensuremath{CurPtr}\xspace}
\newcommand{\curcmd}{\ensuremath{CurCmd}\xspace}
\newcommand{\exgntd}{\ensuremath{ExGntd}\xspace}
\newcommand{\shrset}{\ensuremath{ShrSet}\xspace}

\newcommand{\InvSet}{\ensuremath{\mathcal{I}}\xspace}

\newcommand{\InvConcSet}{\ensuremath{\{1,2\}}\xspace}

\newcommand{\RulesetSingle}[3]{ {\tt Rule} \textit{\textbf{#1}}

\hspace*{0.5cm} {\tt #2}

\hspace*{0.25cm} $\rightarrow$ 

\hspace*{0.5cm} {{\tt #3}} 

\textit{\textbf{End;}}}

\newcommand{\Ruleset}[4]{ {\tt #1} {\tt do}  {\tt Rule} \textit{\textbf{#2}}

\hspace*{0.2cm} {\tt #3 }

\hspace*{0.05cm} $\rightarrow$ 

\hspace*{0.2cm} {\tt #4 }

\textit{\textbf{End;}}}

\newcommand\nop[1]{}
\newcommand{\Other}{{\em Other}\xspace}

\newcommand{\true}{\emph{true}\xspace}
\newcommand{\false}{\emph{false}\xspace}

\newcommand{\Protocol}{\ensuremath{\mathcal{P}}\xspace}

\newcommand{\ie}{{\em i.e.}, }
\newcommand{\etal}{{\em et al.\ }}

\newcommand{\IndexSet}{\ensuremath{\mathbb{N}_N}}
\newcommand{\Enabled}{\emph{Enabled}\xspace}
\newcommand{\g}{\ensuremath{\widehat{en}}}
\newcommand{\ptr}{\ensuremath{\hat{w}}\xspace}

\newcommand{\mdeadlock}{s-deadlock\xspace}
\newcommand{\MDeadlock}{S-deadlock\xspace}
\newcommand{\ISet}[1]{{\ensuremath{In^{#1}}}}
\newcommand{\AssertSet}{\ensuremath{{\mathcal{A}}}\xspace}
\newcommand{\Inv}[1]{{\sc #1}\xspace}
\newcommand{\partialorder}[1]{\ensuremath{{\prec}_{#1}}}
\newcommand{\precon}[3]{\ensuremath{{#2}(#3).p_{#1}}}
\newcommand{\AgentRuleSet}{\ensuremath{\mathcal{R}}}
\newcommand{\flow}{\ensuremath{\mathcal{F}}\xspace}

\newcommand{\revised}[1] {}
\newcommand{\TechnicalVersion}[1] {#1}
\newcommand{\CameraReady}[1] {}

\usepackage{multirow}

\begin{document}


\title{Using Flow Specifications of Parameterized Cache Coherence
  Protocols for Verifying Deadlock Freedom}

\author{Divjyot Sethi$^{1}$ \and Muralidhar Talupur$^{2}$ \and  Sharad Malik$^{1}$}
\institute{
Princeton University\\
\and
Strategic CAD Labs,
Intel Corporation}

\maketitle

\begin{abstract}
We consider the problem of verifying deadlock freedom for symmetric
cache coherence protocols. While there are multiple definitions of
deadlock in the literature, we focus on a specific form of deadlock
which is useful for the cache coherence protocol domain and consistent
with the internal definition of deadlock in the Murphi model checker:
we refer to this deadlock as a {\em system-wide deadlock
  (\mdeadlock)}. In \mdeadlock, the entire system gets blocked and is
unable to make any transition.  Cache coherence protocols consist of N
symmetric cache agents, where N is an unbounded parameter; thus the
verification of \mdeadlock freedom is naturally a parameterized
verification problem.

\nop{
We verify the absence of maximal deadlock (\mdeadlock) --- a deadlock
situation in which the entire system gets stuck --- for symmetric
cache coherence protocols. Cache coherence protocols are parameterized
in the number of agents: they consist of N symmetric cache agents,
where N is an unbounded parameter, interacting through passing
messages. An \mdeadlock error witness involves all of the arbitrarily
large number of agents of the protocol getting stuck and unable to
make any transition. Thus verification of \mdeadlock freedom is
naturally a parameterized verification problem.
}

Parametrized verification techniques work by using sound abstractions
to reduce the unbounded model to a bounded model.
Efficient abstractions which work well for industrial scale protocols
typically bound the model by replacing the state of most of the agents
by an abstract environment, while keeping just one or two agents as
is. However, leveraging such efficient abstractions becomes a
challenge for \mdeadlock: a violation of \mdeadlock is a state in
which the transitions of all of the unbounded number of agents cannot
occur and so a simple abstraction like the one above will not preserve
this violation. Authors of a prior paper, in fact, proposed using a
combination of over and under abstractions for verifying such
properties.  While quite promising for a large class of deadlock
errors, simultaneously tuning over and under abstractions can become
complex.

\nop{
for most of the
parameterized techniques which typically work by discarding some of the
agent state and adding extra transitions in order to use abstractions. The added
transitions make it difficult to preserve this violation.
}

In this work we address this challenge by presenting a technique which
leverages high-level information about the protocols, in the form of
message sequence diagrams referred to as {\em flows}, for constructing
invariants that are collectively stronger than \mdeadlock. Further,
violations of these invariants can involve only one or two interacting
agents: thus they can be verified using efficient abstractions like
the ones described above. We show how such invariants for the German
and Flash protocols can be successfully derived using our technique
and then be verified.

\nop{ Verification of deadlock freedom for message passing protocols
  is an important practical problem. Message passing protocols are
  parameterized in the number of agents: they consist of $N$ agents
  (where $N$ is an unbounded parameter) which interact through passing
  messages. Their verification, then, naturally lends itself to
  parameterized verification techniques.

Verifying deadlock style properties using parameterized techniques is
hard because it is difficult to construct a suitable
abstraction. Abstractions typically add more behaviors to the model
(and thus are over-abstractions). A deadlock in general may involve an
arbitrarily large number of agents stuck in a classic deadlock cycle:
any over-abstraction needs to ensure that the behaviors it adds does
not break any deadlock cycle - this is hard for most scalable
abstractions. Authors in a prior paper had to resort to manually tuned
mixed abstractions (i.e. combination of under and over abstractions)
to verify deadlock freedom.

In this work we present a technique for constructing invariants which
collectively are stronger than the deadlock-style properties under
check. Further, these invariants typically have one or two index
specification and thus are amenable to the state of the art
parameterized techniques.  The construction of these invariants
leverages protocol designer's knowledge and results in very efficient
abstractions. We verify German and Flash protocols using these
techniques.
}

\nop{
Verification of deadlock freedom for message passing protocols
is considered an important problem. 
State of the art techniques for verifying industrial scale protocols
for an arbitrary number of agents (for e.g. caches),
i.e. parameterized verification techniques, are only suitable for
verifying properties like mutual exclusion (i.e. two caches are not in
exclusive state at the same time) which can be specified on a small
number, say $c$, of caches. Further, violations of such properties
involve the state of $c$ caches only.  These enable efficient
abstractions like data type reduction, which keep $c$ caches as is and
replace the effect of others with an environment model. A deadlock
witness, however, may have an arbitrarily large number of caches
involved in a resource dependence cycle. This precludes specifications
using a small number of caches and thus scalable abstractions. Thus
full formal verification of deadlock is typically considered a hard
classical problem.

In this work we verify protocols by utilizing high level information
about the set of protocol messages used for processing high level
requests---these are referred to as {\em flows} and are readily
available in industrial documents. In particular, we show how
properties on these flows can be used to prove deadlock freedom.
\nop{These properties are synthesized iteratively by involving user input
to point out which flow (request) is being processed by the protocol
at a particular time: some flow being processed at all times implies
no deadlock.  are synthesized from flows by the user
}
}
\end{abstract}

\section{Introduction}
We consider the problem of verifying deadlock freedom for symmetric
cache coherence protocols.  Consider a cache coherence protocol
\Protocol(N) where the parameter \(N\) represents an unbounded number
of cache agents. The protocol implements requests sent by the agents
using messages exchanged in the protocol. For a protocol designer, the
main property of interest is the request-response property, i.e.,
every request from an agent eventually gets a response. Since this
property is a liveness property which is hard for existing model
checking tools, designers resort to identifying causes for response
property failure, such as deadlock-style failures, and verify against
them.

The literature is abundant with various definitions of
deadlock~\cite{DeadlockClassic, DeadlockBingham}. We focus on deadlock
errors in which the entire protocol gets blocked, i.e., no agent of
the protocol can make any transition. We refer to such an error as a
{\em system-wide deadlock (\mdeadlock)}. If we model each transition
$\tau$ of the protocol to have a guard $\tau.g$, which is $false$ if
the transition is not enabled, the \mdeadlock error occurs if the
guards of all the transitions are \false, i.e., \(\bigwedge_{\tau}
\lnot (\tau.g)\) is \true. This kind of failure, while weaker than
other broader classes of deadlock failures, is commonly observed in
industrial computer system designs and is consistent with the internal
definition for deadlock used by the Murphi model checker as
well~\cite{Murphi}. This class of deadlocks is well motivated for
parameterized cache coherence protocols as these use a centralized
synchronization mechanism (e.g. a directory) and thus any deadlock
results in the directory getting blocked. It is highly likely that
such a deadlock in the shared directory will end up involving all of
the agents of the protocol getting blocked, i.e., unable to make any
transition.

Since an \mdeadlock error involves all of the unbounded number of
agents getting blocked and unable to make any transition, verification
of \mdeadlock freedom naturally is a parameterized verification
problem. Parameterized verification techniques work by using sound
abstractions to reduce the unbounded model to a finite bounded model
that preserves the property of interest. These abstractions typically
tend to be simple over-abstractions such as {\em data-type
  reduction}~\cite{ConcDSDataTypeRed}. This abstraction keeps a small
number of agents ($1$ or $2$) as is and replaces all the other agents
with an abstract environment. Such abstractions along with
parameterized techniques like the CMP (CoMPositional)
method~\cite{CMP} have had considerable success in verifying key
safety properties like mutual exclusion and data integrity even for
industrial scale protocols~\cite{CMP, ParamVerifMuraliIndust,
  ParamVerMurali}.

\subsection{Challenge in Verifying \MDeadlock}
While parameterized techniques are successful for safety properties
such as mutual exclusion and data integrity, the application of such
abstractions for parameterized verification of properties such as
\mdeadlock is hard. The key challenge arises from the fact that an
\mdeadlock violation is a state in which all the guards are \false,
i.e., when \(\bigwedge_{\tau} \lnot (\tau.g)\) holds; simple
over-abstractions such as data-type reduction will easily mask this
violation due to the discarded state of agents other than $1$ and $2$
and the extra transitions of the environment.

One approach to address the above issue is to use a combination of
over and under abstractions (i.e., a {\em mixed} abstraction) instead
of data-type reduction, as described in a prior deadlock verification
work~\cite{DeadlockBingham}. While promising for verifying a large
class of deadlock errors, the use of mixed abstraction requires
reasoning about over and under abstraction simultaneously and easily
becomes fairly complex.

In this paper we take a different approach. We show how high-level
information about the protocols, in the form of message sequence
diagrams referred to as {\em flows}, can be leveraged to construct
invariants which are collectively stronger than the \mdeadlock freedom
property. These invariants are amenable to efficient abstractions like
data-type reduction which have been used in the past for verifying
industrial scale protocols.

\subsection{Leveraging Flows for Deadlock Freedom}
Cache coherence protocols implement high-level requests for read
(termed \Shared) or write (termed \Exclusive) access from cache
agents, or for invalidating access rights (termed \Invalidate) of some
agent from the central directory. The implementation of these requests
is done by using a set of transitions which should occur in a specific
protocol order. This ordering information is present in diagrams
referred to as {\em message flows} (or {\em flows} for brevity). These
flows are readily available in industrial documents in the form of
message sequence charts and tables~\cite{ParamVerMurali}.

\begin{figure}
\centering
\hspace*{- 0.75cm}
\subfloat[\Exclusive flow]{
\label{figGermanFlowsExcl}
  \includegraphics[scale=.55]{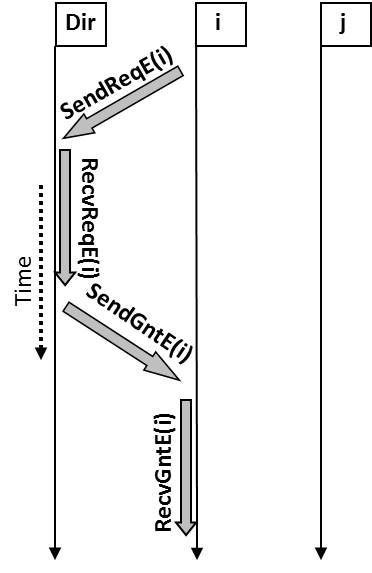}
}
\hspace*{1.75cm}
\subfloat[\Invalidate flow]
{
\label{figGermanFlowsInv}
  \includegraphics[scale=.55]{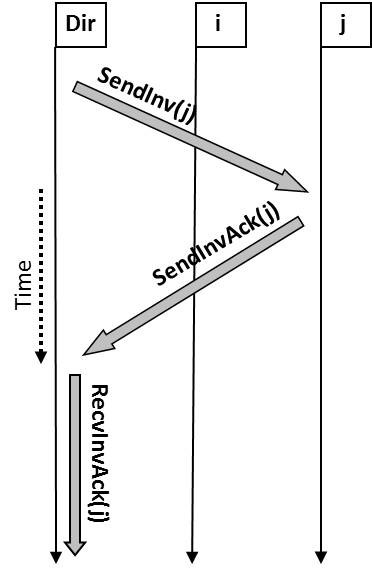}
}
\caption{Flows for the German protocol.}
\label{figGermanFlows}
\end{figure}

\Fig{figGermanFlows} shows two of the flows for the German cache
coherence protocol describing the processing of the \Exclusive and
\Invalidate requests. Each figure has a directory $Dir$, and two
agents $i$ and $j$. The downward vertical direction indicates the
passage of time. The \Exclusive request is sent by the cache agent $i$
to the directory $Dir$ to request a write access. The \Exclusive flow
in \SubFig{figGermanFlowsExcl} describes the temporal sequence of
transitions which occur in the implementation in order to process this
request: each message is a transition of the protocol. The message
$SendReqE(i)$ is sent by the agent $i$ to $Dir$ which receives this
message by executing the transition $RecvReqE(i)$.  Next, if the
directory is able to grant \Exclusive access, it sends the message
$SendGntE(i)$ to agent $i$ which receives this grant by executing
$RecvGntE(i)$. However, in case the directory is unable to send the
grant since another agent $j$ has access to the cache line, the
directory sends a request to invalidate the access rights of $j$. The
temporal sequence of transitions which occur in the implementation in
this case are shown in the \Invalidate flow in
\SubFig{figGermanFlowsInv}. This flow proceeds by the directory sending
the $SendInv(j)$ message, the agent $j$ sending the acknowledgment
message $SendInvAck(j)$, and the directory receiving it by executing
$RecvInvAck(j)$ transition.

\nop{
Next, it may be noted that the message $SendGntE(i)$ is sent only if
no other agent has \Shared access to the cache line. In case another
agent, say $j$, has access to the cache line, the directory uses the
\Invalidate(j) flow to first invalidate $j$. This flow proceeds by the
directory sending the invalidate ($SendInv(j)$), the agent $j$ sending
the acknowledgment ($SendInvAck(i)$), and the directory receiving it
($RecvInvAck(i)$).
}
\nop{
\subsubsection{Deadlock Situation:} Now consider the case when in a buggy protocol
the agent $j$ does not send the acknowledgment $SendInvAck(i)$ due to
a coding error. In this case, the directory is blocked and waiting to
receive $SendInvAck(i)$. The blockage of directory leads to all the
$N$ agents in the protocol getting blocked as well. Thus no transition
in the protocol is enabled. In an RTL implementation of such a
protocol, one would assume the directory to timeout if the
$SendInvAck(i)$ is not received: even in that case, this kind of a
high-level bug would seriously impact the performance of the protocol.
}

\subsubsection{Freedom from \MDeadlock}
At a high-level, our method tries to exploit the fact that if the
protocol is \mdeadlock free, when none of the transitions of an agent
are enabled, another agent can be identified which must have a
transition enabled. This identification leverages the key insight that
in any state of the protocol, if all the transitions of some agent,
say $a_1$, cannot occur, then, some flow of that agent must be blocked
since it depends on another flow of another agent, say $a_2$, to
finish. Then, there are two possibilities: (1) the agent $a_2$ is
enabled, in which case the state is not an \mdeadlock state, or (2)
the agent $a_2$ is blocked as well, in which case it depends on
another agent $a_3$. If this dependence chain is acyclic, with the
final agent in the chain enabled, the protocol is \mdeadlock
free. However, if the final agent is not enabled, or if the dependence
chain has a cycle, the protocol may either have an \mdeadlock error or
there may be an error in the flow diagrams used.

As an example, for the German protocol, if the \Exclusive flow of
agent $i$ is blocked since the transition $SendGntE(i)$ cannot occur,
it is waiting for $j$ to get invalidated. In the protocol, at least
some transition of the \Invalidate flow on agent $j$ can occur. This
enables proving freedom from \mdeadlock for the protocol.

Using the above insight, by analyzing the dependence between blocked
agents, our method is able to point to an agent which must have at
least one transition enabled in every reachable state of the
protocol. Specifically, our method enables the derivation of a set of
invariants $\InvSet$ which collectively partition the reachable state
of the protocol. Each invariant then points to the agents which must
have at least one transition enabled when the protocol is in a state
belonging to its partition. These invariants are derived in a loop by
iteratively model checking them on a protocol model with $c$ agents,
where $c$ is heuristically chosen as discussed in
\Sec{sec:SpecMethod}.

\subsubsection{Verifying for an Unbounded Number of Agents}
Once the invariants in $\InvSet$ are derived, they hold for a model
with $c$ agents. These invariants use just one index (i.e., they are
of the form $\forall i: \phi(i)$) and thus, they can be verified for
an unbounded number of agents by using efficient parameterized
verification techniques such as data-type reduction along with the CMP
(CoMPositional) method~\cite{CMP}. This technique has previously been
successful for verifying mutual exclusion for industrial
protocols~\cite{ParamVerifMuraliIndust}. We note that our approach is
not limited to the CMP method: the invariants derived may be verified
by using any parameterized safety verification
technique~\cite{InvisibleInvariants, CounterAbstraction,
  IndexedPredicates, ProtocolFlashAmitGoel}.

\nop{
Our method takes a Murphi model of the protocol along with an initial
broad guess for $\InvSet$ as input. Then, $\InvSet$ is iteratively
refined by model checking it on a protocol model with $c$ agents in a
loop. This refinement is done by user-assisted analysis of the
returned counterexample and splitting the failing invariant in
$\InvSet$ by leveraging flow information. Once the invariants in
$\InvSet$ pass for a model with $c$ agents, they are verified for an
unbounded number of agents using data-type reduction, along with the
CMP (CoMPositional) method~\cite{CMP}, a previously known technique
highly successful for verifying mutual exclusion for industrial
protocols~\cite{ParamVerifMuraliIndust}. \Fig{figExperimentalSetup}
shows the setup for our method, and highlights the key contribution of
this paper.

It may be noted that our approach is not limited to the CMP method:
the invariants derived may be verified using any of the numerous
parameterized safety verification
techniques~\cite{InvisibleInvariants, CounterAbstraction,
  IndexedPredicates}.
}

\nop{
Our method tries to partition the state of the protocol such that in
each partition it can point to an agent $i$ and claim that one of its
transitions must be enabled. This is done by starting with an initial
guess of state partition for which agent $i$ must have an enabled
transition.  In case the agent does not have any transition enabled,
typically there is a particular flow of the agent which is 'stuck'
because the protocol is processing a flow of another agent $j$. This
information is used to refine the partition. This way, we derive
invariants of the form: \(pred \rightarrow \phi(i)\), where $pred$ is
a predicate on protocol state and $\phi(i)$ is a disjunction of
transitions for agent $i$. Invariants of this form are much easier to
check using existing parameterized verification techniques.
}

\subsection{Key Contributions}
\begin{figure}
\centering
\hspace*{-0.5cm}  \includegraphics[scale=.55]{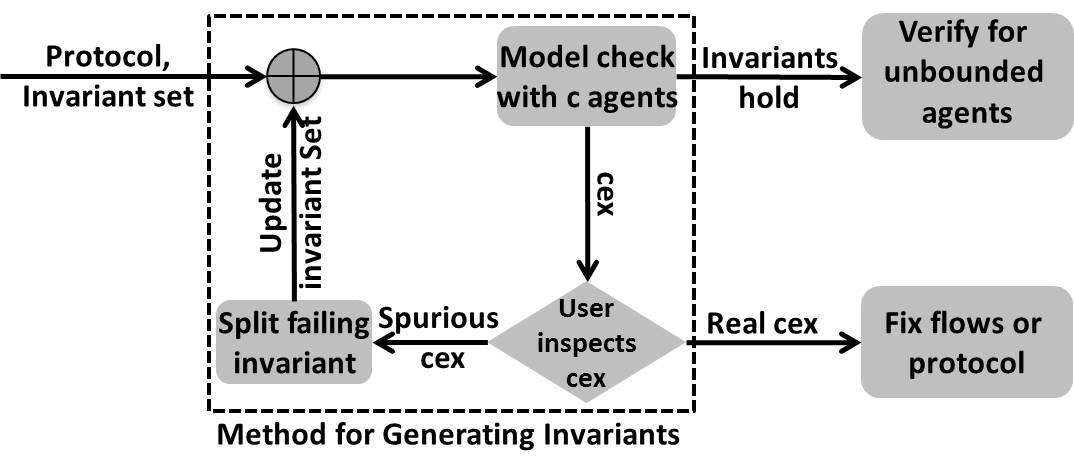}
\caption{Experimental Flow}
\label{figExperimentalSetup}
\end{figure}

Our method proves \mdeadlock freedom for parameterized protocols
(formalized in \Sec{sec:formalModel}). It takes a Murphi model of the
protocol as input. As shown in \Fig{figExperimentalSetup}, first, a
set of invariants $\InvSet$ which collectively imply \mdeadlock
freedom are derived on a model with $c$ agents
(\Sec{sec:SpecMethod}). These invariants are verified for an unbounded
number of agents by using state-of-the-art parameterized verification
techniques (\Sec{sec:CMP}).  We verified Murphi implementations of two
challenging protocols, the German and Flash protocols using our method
(\Sec{sec:Experiments}).

{\em Limitation:} The key limitation of our approach is that the
invariants have to be derived manually by inspecting
counterexamples. This can be automated if additional information about
conflicting flows is available in the flow diagram itself.

\subsection{Relevant Related Work}
\textbf{{\em Deadlock Verification:}} The work closest to ours is by
Bingham \etal~\cite{DeadlockBingham, DeadlockBinghamCAV}. They
formally verify deadlock as a safety property for protocols by
specifying it using user-identified Quiescent states (\ie a state in
which no resources are held): they specify a protocol state to be a
deadlock state if no Quiescent state is reachable from it. They prove
freedom from such a deadlock by using a combination of over and under
abstractions (also referred to as a {\em mixed}
abstraction~\cite{mixedAbstractions}).
Their approach is a promising way to verify deadlock freedom which
scales to protocols like the Flash protocol. However, the required
tuning of both under and over abstractions simultaneously can be
complex. In contrast, we take the flow-based alternative to enable
simpler abstractions like data-type reduction.

Since the ultimate goal of any deadlock verification effort is to
verify the response property (i.e. every high-level request eventually
gets a response), we contrast our work with liveness verification
efforts as well. Among techniques for parameterized verification of
liveness, McMillan has verified liveness properties of the Flash
protocol~\cite{DeadlockFlash, McMillanLiveness}. The proof is manual
and works on the basis of user supplied lemmas and fairness
assumptions. In contrast, our method reduces manual effort by
leveraging information from flows along with the CMP method.
Among automatic approaches for verifying liveness properties, Baukus
\etal verified liveness properties of the German
protocol~\cite{DeadlockWSIS} using a specialized logic called
WSIS. Fang \etal used automatically deduced ranking
functions~\cite{InvisibleRanking} and, in a prior work, counter
abstraction~\cite{CounterAbstraction} to verify liveness
properties. While fully automatic, these approaches tend to exhibit
limited scalability for larger protocols such as Flash, due to the
inherent complexity of the liveness verification problem. In contrast
to these, our approach, while requiring some user guidance, achieves
much greater scalability and enables us to verify the Flash protocol.

\textbf{{\em Parameterized Verification Techniques:}} We note that the
invariants derived using our method can be verified for an unbounded
number of caches by any parameterized safety verification technique,
it is not dependent on the CMP method which we used. Our choice of
using the CMP method was motivated by the fact that it is the only
state-of-the-art method we are aware of which has been used
successfully for verifying protocols like Flash and other industrial
scale protocols. Among other techniques, an important technique is by
Conchon \etal~\cite{ProtocolFlashAmitGoel} which uses a backward
reachability algorithm to automatically prove a simplified version of
the Flash protocol. Next, there are numerous other prior approaches in
literature for parameterized verification of safety properties. The
CMP method falls in the broad category of approaches which use {\em
  compositional reasoning}~\cite{ParamLamportCompositional,
  ParamLamportCompositional1} and {\em abstraction} based techniques
to verify parameterized systems; the literature is abundant with
examples of these~\cite{ParamVerifNamjoshiCompositional,
  DeadlockFlash, InvisibleInvariants, CounterAbstraction,
  IndexedPredicates, DasDill:99,
  ParamVerifEnvironmentAbstractionMurali, ParamVerifEnvAbstrMurali}.
Next, another category of approaches work by computing a {\em cutoff}
bound $k$ and showing that if the verification succeeds for $k$
agents, then the protocol is correct for an arbitrary number of
agents~\cite{ParamAmir1, ParamKahlon, ParamVerifVineetCutoff,
  ParamVerifClarkeCutoff, ParamWahlCutoff,
  ParamAbdullaCutoff}. Finally, there are approaches based on {\em
  regular model checking} which use automata-based algorithms to
verify parameterized systems~\cite{ParamRegularModelCheckingClassic,
  PAAbdullahRegularModelCheckingSurvey,
  ParamRegularModelCheckingClassic1, ParamIteratingTransducers}. To
the best of our knowledge, the CMP method is the state-of-the-art for
protocol verification in contrast to these methods and has been used
to successfully verify larger protocols such as Flash with minimal
manual effort. (Other methods which verify Flash protocol in full
complexity are by Park \etal~\cite{ParkDill:96} and Park
\etal~\cite{DasDill:99}. As described by Talupur
\etal~\cite{ParamVerMurali}, these are significantly manual and take
much more time to finish verification of the Flash protocol compared
to the CMP method.)

\nop{ Finally, the last category of approach thr bounded, some
  examples include Invisible invariants~\cite{InvisibleInvariants},
  Counter abstraction~\cite{CounterAbstraction}, Indexed
  predicates~\cite{IndexedPredicates}, and cutoff based
  techniques~\cite{KahlonManyToFew}: }

\nop{
In contrast to all the methods above, our method doesn't develop any
new model checking algorithms for handling deadlock freedom. It
instead strengthens deadlock freedom to a set of universally
quantified small-indexed safety properties which can be verified using
standard techniques. In a loose sense, our approach to strengthening
deadlock freedom verification is similar in spirit to the approach in
the Xmas framework~\cite{xmasVmcai, DeadlockSayak} used for generating
invariants for proving progress properties.

Finally, among parameterized verification techniques for safety, we chose
to use the CMP (CoMPositional) method~\cite{CMP} for 

, there are numerous
other techniques for verifying safety properties. Some examples
include Invisible Invariants for safety~\cite{InvisibleInvariants},
counter abstraction for safety~\cite{CounterAbstraction}, Indexed
predicates~\cite{IndexedPredicates}. The CMP method is a highly
successful parameterized verification technique, numerous work has
been done in the area of parameterized verification, for
e.g.~\cite{InvisibleInvariants, CounterAbstraction,
  IndexedPredicates}. While some of these classical methods are
potentially applicable to verifying protocols as well, given the
success of the CMP method in verifying industrial cache coherence
protocols~\cite{ParamVerifMuraliIndust}, we believe that our method,
which has a similar flavor to the CMP method, provides a scalable and
effective alternative approach.

}

\section{Protocols, Flows and \MDeadlock Freedom: Background}
\label{sec:formalModel}
\subsection{Preliminaries}

A protocol \Protocol(N) consists of $N$ symmetric cache agents, with
{\it ids} from the set $\IndexSet = \{1,2,3,\ldots,N\}$. We follow our
prior approach~\cite{ParamVerMurali} (which was inspired by the
approach of Kristic~\cite{ParamVerifCMPKristic}) in formalizing cache
coherence protocols.

{\em Index Variables:} The protocol uses index variables quantified
over the set of index values $\IndexSet$. Thus, if $i$ is an index
variable, then $i$ takes values from the domain $\IndexSet$.

{\em State Variables:} The state of the protocol consists of local
variables, and global variables shared between the agents. Each of
these types of variables can either hold values from the Boolean
domain (variables with values from generic finite domains can be
represented as a set of Boolean variables) or pointers which can hold
agent {\it ids}. We represent the Boolean variables in the global
state as $G_B$, and the pointers as $G_P$. The Boolean local state
variables of each agent $i$ are encoded as $L_B[i]$, and the local
pointer variables as $L_P[i]$.  The pointer variables have values from
the domain $\IndexSet \cup \{null\}$, where $null$ represents that the
variable does not hold any index value.

{\em Expressions:} An expression is a, possibly quantified,
propositional formula with atoms $G_B$, $G_P=j$, $L_B[i]$ and
$L_P[i]=j$, where, $i$ and $j$ are index variables.

{\em Assignments:} Assignments are of the form $G_B := b$, or
$G_P:=j$, $L_B[i] := b$ or $L_P[i]:=j$, where, $b$ is a variable with
Boolean value and $i$, $j$ are index variables.

{\em Rules: } Each agent $i$ consists of a set of rules
$rl_1(i),rl_2(i),rl_3(i),\ldots,rl_k(i)$. Each rule $rl_j(i)$ can be
written as: \(rl_j(i): rl_j(i).\rho \rightarrow rl_j(i).a,\) where,
$rl_j(i)$ is the rule name, $rl_j(i).\rho$, the guard, is an
expression, and $rl_j(i).a$ is a list of assignments, such that these
assignments are restricted to only update the global variables or the
local variables of agent $i$.  The local variables and rules for all
agents $i$ are symmetric.

{\em Protocol:} The above defined variables and rules naturally induce
a state transition system. A {\em protocol}, then, is a state
transition system $(S,\Theta,T)$, where $S$ is the set of protocol
states, $\Theta \subseteq S$ is the set of initial states, and $T
\subseteq S \times S$ is the transition relation. Each protocol state
$s \in S$ is a valuation of the variables $G_B$, $G_P$, and $L_B[i]$,
$L_P[i]$ for each agent $i$. There exists a transition
$\tau(i_v)=(s,s'), (s,s')\in T$ from state $s$ to $s'$ if there is a
rule $rl_j(i)$ and value of index variable $i=i_v$,
s.t. $rl_j(i_v).\rho$ holds in $s$, and $s'$ is obtained by applying
$rl_j(i_v).a$ to $s$. In state $s$, we say that the rule $rl_j(i)$ is
{\em enabled} for agent with {\it id} $i_v$ if the guard
$rl_j(i_v).\rho$ is \true.  When the enabled rule is executed, its
action is applied to update the state and we say that the rule
$rl_j(i)$ has {\em fired} for agent $i_v$. The action is applied
atomically to update the state, thus the transitions of the protocol
have interleaving semantics. Finally, we define an execution {\em
  trace} of the protocol as a series of transitions where each
transition is a fired rule. Thus, a trace can be represented by a
series ($rl_a(i_0)$, $rl_b(i_1)$, \ldots, $rl_s(i_k)$), where the
transition $rl_m(i_n)$ is the rule $rl_m$ fired for the agent with \id
$i_n$.

\nop{
{\em Transitions:} The transitions of the protocol are indexed over
$i$, \(\tau_1(i),\tau_2(i),\tau_3(i) \ldots \tau_k(i)\). In this
paper, we express the state transitions of the protocol using {\em
  rules} (we use the words rule and transition interchangeably). A
rule $rl(i)$ is a guard-action pair which is written as $rl(i).\rho
\Rightarrow rl(i).a$, where the guard $rl(i).\rho$ (also referred to
as $\rho(i)$ for brevity) is a Boolean expression on variables in
$V_L[i]$ and $V_S$, and the action $rl(i).a$ (also referred to as
$a(i)$ for brevity) is an update to variables in $V_L[i]$ and
$V_S$. We say that a rule is {\em enabled} in a state $s$ if its guard
$\rho$ evaluates to $true$ in $s$, else the rule is {\em disabled}. An
enabled rule is said to have fired when it applies its action $a$ to
update the state. When a rule fires, the corresponding transition
occurs.
}

{\bf \MDeadlock Definition}
We define a protocol state $s$ to be an {\em \mdeadlock state} if no
rule in that state is enabled. Then, a protocol is \mdeadlock free if
in all states, there exists at least one rule which is enabled. This
can be expressed as the invariant: \(\bigvee_{i} \bigvee_{j}
rl_j(i).\rho\), i.e., the protocol is \mdeadlock free if the
disjunction of the guards of all the rules of all the agents is \true
for all the reachable states.



{\bf Flows} Flows describe the basic organization of rules for
implementing the high-level requests in a protocol (for example a
request for \Exclusive access or an \Invalidate). We model a flow as a
set of rules $\flow(i)$ of the form $\{$$rl_a(i), rl_b(i), rl_c(i),
\ldots, rl_n(i)$$\}$ which accomplish a high-level request of agent
$i$.\footnote{For ease of exposition we assume that the guard and
  action of a rule are over the variables of a single agent. Thus, a
  flow containing such rules also involves a single agent. In general,
  a rule and thus a flow can involve a larger but fixed number of
  interacting agents as well. Our approach can be easily generalized
  to that case.} The rules in a flow are partially ordered, with the
partial order relation denoted as \partialorder{\flow(i)}. For
example, in the \Exclusive flow in \SubFig{figGermanFlowsExcl}, the rules
(arrows) are totally ordered along the downward direction. Thus
$SendReqE(i)$ $\partialorder{\flow_E(i)}$ $RecvReqE(i)$, where
$\flow_E$ denotes the set of rules for \Exclusive flow. For every rule
$rl_k(i)$ in the flow $\flow(i)$, the partial order naturally induces
the following {\em precondition}: for the rule $rl_k(i)$ to fire, all
the rules preceding that rule in the partial order of the flow
$\flow(i)$ must have already been fired. This precondition is denoted
by $\precon{\flow(i)}{rl_k}{i}$ and, formally, can be written as:
\[\precon{\flow(i)}{rl_k}{i} = \forall j:\big(\{(rl_j(i) \in \flow(i)) \land
(rl_j(i) \partialorder{\flow(i)} rl_k(i))\} \Rightarrow (rl_j(i).fired
= true)\big),\] where $rl_j(i).fired$ is an auxiliary variable which
is initially set to \false when the flow $\flow(i)$ starts and is set
to \true when the rule $rl_j(i)$ has fired for that flow.

\nop{
Then,
given a rule $rl_k(i)$, the flow $fl(i)$ naturally induces a {\em
  precondition} that for the rule to be enabled, all the rules
$rl_a(i), rl_b(i), rl_c(i) ,\ldots, rl_{k-1}(i)$ must have already
fired. Formally, we denote this by $rl_k(i).pre^{fl}$ which holds on a
state if the rules preceding $rl_k(i)$ have all fired.
}

\revised{This Para} Designs of protocols are presented in industrial
documents as a set of flows $\flow_1(i)$, $\flow_2(i)$, $\flow_3(i)$,
$\ldots$, $\flow_k(i)$. In order to process a high-level request, a
protocol may use a combination of these flows, e.g. in order to
execute a request for \Exclusive access the German protocol uses the
\Exclusive and \Invalidate flows. Each flow in a protocol represents
an execution scenario of the protocol for processing some high-level
request. Thus many of the flows of a protocol tend to exhibit a lot of
similarity as they are different execution scenarios of the same
high-level request.  This makes them fairly easy to understand. In
\Sec{sec:SpecMethod}, we show how a set of invariants collectively
implying \mdeadlock freedom can be derived from these flows.

{\bf Some definitions:}
We define the union of all the flows of agent $i$ by
$\AgentRuleSet(i)$, i.e., $\AgentRuleSet(i)=\bigcup_{k}\flow_k(i)$.
Next, we define the operator $\g$ which is \true for a set of rules,
if at least one rule in the set is enabled, else it is \false. Thus,
for example, $\g(\AgentRuleSet(i))$ holds if at least one of the rules
in $\AgentRuleSet(i)$ is enabled. In this case, we say that the agent
$i$ is enabled. Similarly, we say that a flow $\flow(i)$ is enabled if
at least one of its rules is enabled, i.e., $\g(\flow(i))$ holds. In
case a flow $\flow(i)$ is not enabled, we say that it is {\em blocked}
on some rule $rl_j(i) \in \flow(i)$ if the precondition of the rule
$\precon{\flow(i)}{rl_j}{i}$ holds but the guard of the rule
$rl_j(i).\rho$ is \false. 

\subsection{German Protocol Implementation}
\nop{The German protocol implements 3 high-level requests: (1) an
\Exclusive request initiated by an agent for getting a read or a write
permission (agent state with read and write permission is denoted by
$E$), (2) a \Shared request for getting a read only permission (agent
state denoted by $S$), and (3) an \Invalidate request initiated by the
directory to take away all the permissions from an agent (agent state
with no permissions is denoted by $I$).
}


The German protocol consists of agents such that each agent can have
\Exclusive ($E$), \Shared ($S$) or \Invalid ($I$) access to a cache
line, as stored in the variable $Cache[i].State$. An agent $i$
requests these access rights by sending messages on a channel
$ReqChannel[i]$ to a shared directory which sends corresponding grants
along the channel $GntChannel[i]$. The directory is modeled as a set
of global variables which serves one agent at a time: it stores the
{\it id} of the agent being served in the variable $CurPtr$. It also
stores the nature of the request in the variable $CurCmd$ with values
in $\{ReqE,ReqS,Empty\}$, where $ReqE$ represents a request for
\Exclusive access, $ReqS$ for \Shared and $Empty$ for no
request. Finally, the directory tracks if \Exclusive access is granted
to some agent or not using the variable $ExGntd$: it is \true if
access is granted and \false otherwise. \CameraReady{A
  simplified version of the code for the \Exclusive request is shown
  in \Fig{Fig:ExclusiveCode}.} \TechnicalVersion{A
  simplified version of the code for the \Exclusive request is shown
  in \Fig{Fig:ExclusiveCode}, with the original Murphi
  implementation~\cite{CMP} presented in
  Appendix~\ref{AppendixGermaCode}.}

\begin{figure}[t]
\centering
    \begin{minipage}{\textwidth}
        \Ruleset{$\forall$ i : \IndexSet;}{SendReqE(i)}{ReqChannel[i].cmd=Empty $\land$ 

\hspace*{.45cm}(Cache[i].State=I $\lor$ Cache[i].State=S)}{ReqChannel[i].cmd := ReqE;}

\vspace*{.25cm} \Ruleset{$\forall$ i :
  \IndexSet;}{RecvReqE(i)}{ReqChannel[i].cmd=ReqE $\land$
  CurCmd=Empty}{CurCmd := ReqE; CurPtr := i;

\hspace*{.35cm} ReqChannel[i].cmd := Empty;}        

\vspace*{.25cm}
        \Ruleset{$\forall$ i : \IndexSet;}{SendGntE(i)}{CurCmd=ReqE $\land$ CurPtr=i $\land$ 

\hspace*{.45cm} GntChannel[i]=Empty $\land$ Exgntd=false 

\hspace*{.45cm}$\land$ ShrSet=\{\}}{GntChannel[i] := GntE; ShrSet := \{i\}; 

\hspace*{.35cm}ExGntd := true; CurCmd := Empty; 

\hspace*{.35cm}CurPtr := NULL;}  

\vspace*{.25cm}
        \Ruleset{$\forall$ i : \IndexSet;}{RecvGntE(i)}{GntChannel[i]=GntE}{Cache[i].State := E; GntChannel[i] := Empty;} 
    \end{minipage}
\caption{Implementation of the \Exclusive Request.}
\label{Fig:ExclusiveCode}
\end{figure}

\nop{
\begin{figure}
\centering
    \begin{minipage}{\textwidth}
        \Ruleset{$\forall i : \IndexSet;$}{SendInv}{$(invChannel[i].cmd =
          Empty) \land (i \in ShrSet) \land (InvSent[i] = false) \land $

$((CurCmd = ReqE) \lor ((CurCmd = ReqS) \land (ExGntd = true)))$}{$invChannel[i].cmd = Inv;$ $InvSent[i] = true;$}

        \Ruleset{$\forall i : \IndexSet;$}{SendInvAck}{$(invAckChannel[i].cmd =
          Empty) \land (invChannel[i].cmd = Inv)$}{$invChannel[i].cmd = Empty;$ $invAckChannel[i].cmd = invAck;$ $writeback$ $data;$}        

        \Ruleset{$\forall i : \IndexSet;$}{RecvInvAck}{$(invAckChannel[i].cmd = InvAck) \land (CurCmd != Empty)$}{$invAckChannel[i] = Empty;$ $ShrSet = ShrSet
\setminus
\{i\};$ $if(ExGntd)$ $then$ $ExGntd := false;$ $Update$ $memory;$ $endif;$}  
    \end{minipage}
\caption{Invalidate Flow}
\label{Fig:InvalidateFlow}
\end{figure}
}

In processing the \Exclusive request, before sending the grant
$SendGntE(i)$, the directory checks if there are any sharers of the
cache line (by checking $ShrSet$ = $\{\}$). If there are sharers, the
\Invalidate flow is invoked for each agent in $ShrSet$.  Upon
invalidation of all the agents in $ShrSet$, the $ShrSet$ becomes empty
and so the $SendGntE(i)$ rule becomes enabled for execution. We show
the code for the $SendInv(i)$ rule below.

\Ruleset{$\forall$ i : \IndexSet;}{SendInv(i)}{InvChannel[i].cmd =
  Empty $\land$ i $\in$ ShrSet $\land$ 

\hspace*{.4cm}  ((CurCmd = ReqE) $\lor$ (CurCmd = ReqS $\land$ ExGntd =
  true))}{InvChannel[i].cmd := Invalidate;}

We note a condition $Inv\_Cond$, which must be \true for invoking the
\Invalidate flow and can be identified from the guard of $SendInv(i)$;
\(Inv\_Cond: \big(\big((CurCmd = ReqE) \lor ((CurCmd=ReqS) \land
(ExGntd=true))\big) \land (ShrSet \neq \{\})\big) \).





\nop{
The German protocol models a directory as a set of global variables.
The directory stores the index value of the agent whose request is
currently being served in a global variable $CurPtr$ with values in
$\{\{null\} \cup [1,N]\}$. $CurPtr=i$ if request of agent $i$ is being
served and $CurPtr=null$ if no request is being served. Next,
information about the current request being served by the directory is
encoded in the variable $CurCmd$ with values in $\{ReqE,ReqS,Empty\}$,
where $ReqE$ represents a request for \Exclusive access, $ReqS$ for
\Shared access and $Empty$ for no request. The directory also
maintains a list of agents which share a particular cache line,
denoted by $ShrSet$. Finally, the directory tracks if \Exclusive
access is granted to some agent or not using the Boolean variable
$ExGntd$: it is $true$ if access is granted and $false$ otherwise.

Each agent $i$ stores its current status in local variable
$Cache[i].State$ with values in ${I,S,E}$, where $I$, $S$ and $E$
stand for $Invalid$, $Shared$ and $Exclusive$, respectively. Next, the
agents and directory communicate with each other through channels: the
implementation of the German protocol uses separate channels for
separate requests. Agent $i$ sends requests for \Shared or \Exclusive
access to the directory on $ReqChannel[i]$. The directory in turn
sends grants to agent $i$ using the channel $GntChannel[i]$.  Finally,
the directory sends \Invalidate to agents through the channel
$InvChannel[i]$: in order to track that an \Invalidate request is sent
to a particular agent $i$, it keeps a variable $InvSent[i]$ which is
$true$ if the request has been sent and $false$ otherwise.

\paragraph{Code for \Exclusive request}
\Fig{Fig:ExclusiveCode} shows the relevant code details\footnote{For
  increasing clarity, compared to the original code presented in
  Appendix~\ref{AppendixGermaCode}, some variable names have been
  changed as well as some details not relevant to the discussion have
  been skipped. Note that variable names start with capital letters
  for the sake of consistency with the original code.} for the
implementations of the $SendReqE(i)$, $RecvReqE(i)$, $SendGntE(i)$ and
$RecvGntE(i)$ rules of the \Exclusive flow (flow shown in
\Fig{figGermanFlowsExcl}). Notice that the guard of the rule
$SendGntE(i)$ checks if there are no sharers of the cache line
($ShrSet$ = $\{\}$). Rest of the guards and actions of the rules in
the figure are self-explanatory.

\paragraph{Code for \Invalidate request}
For the implementation of the \Invalidate request (flow shown in
\Fig{figGermanFlowsInv}), for the sake of brevity, we only provide the
code for the rule $SendInv(i)$ which implements an \Invalidate request
initiated by the directory for a sharer:

\vspace{.3cm}
\begin{minipage}{.3\textwidth}
\footnotesize
\Ruleset{$\forall$ i : \IndexSet;}{SendInv(i)}{InvChannel[i].cmd =
  Empty $\land$ i $\in$ ShrSet $\land$ InvSent[i] = false 

\hspace*{.4cm}  $\land $ ((CurCmd = ReqE) $\lor$ (CurCmd = ReqS $\land$ ExGntd =
  true))}{InvChannel[i].cmd = Inv; InvSent[i] = true;}
\end{minipage} \vspace{.2cm}

The guard of this rule checks the following two key conditions: (1)
first, it checks if the current request for \Shared or \Exclusive
access requires the directory to initiate an \Invalidate request to
the agents sharing the cache line. This is true when the current
request is either for an \Exclusive access or, is for a \Shared access
with another agent already having an \Exclusive access. This is stated
as the condition: \(Inv\_Cond: CurCmd = ReqE \lor (CurCmd=ReqS \land
ExGntd=true).\) (2) And second, it checks whether there is an agent
$i$ which needs to be invalidated, by checking $(i \in ShrSet)$.
}
\nop{
This rule checks if the channel for sending invalidates
($invChannel[i]$) is empty, that $i$ is in the set of sharers
($ShrSet$), that invalidate has not yet been sent (tracked using
$invSet[i] = false$) and that the current request is either exclusive
or it is for shared but some other cache has exclusive access
(i.e. $((CurCmd = ReqE)$ $\lor$ $((CurCmd = ReqS)$ $\land$ $(ExGntd =
true)))$, denoted by $invCond$).
}

\nop{

The protocol code snippets implementing the Exclusive and Invalidate
requests for the German protocol are shown in \Fig{Fig:ExclusiveFlow}
and \Fig{Fig:InvalidateFlow} respectively\footnote{The complete code
  is in the appendix.}.

\subsubsection{Flows}
The German protocol consists of three flows: \Exclusive flow, \Shared
flow and \Invalidate flow. The \Exclusive flow processes a request for
write access, \Shared for read access and the \Invalidate flow for
invalidating some cache. The \Invalidate flow is a sub-flow of \Shared
and \Exclusive flows. These flows are detailed as follows:

\begin{itemize}
\item{\Exclusive:} {\bf ReqExclusive(i)}: $SendReqE(i)$, $RecvReqE(i)$,
  $SendGntE(i)$, $RecvGntE(i)$.

\item{\Shared:} {\bf ReqShared(i)}: $SendReqS(i)$, $RecvReqS(i)$,
$SendGntS(i)$, $RecvGntS(i)$.

\item{\Invalidate:} {\bf SendInval(i)}: $SendInv(i)$, $SendInvAck(i)$,
  $RecvInvAck(i)$.
\end{itemize}

\subsubsection{Protocol Code}
The code snippets for the Exclusive and Invalidate flows of the German
protocol is presented in Murphi language~\cite{} in
\Fig{Fig:ExclusiveFlow} and \Fig{Fig:InvalidateFlow}
respectively\footnote{The complete code is in the appendix.}.  These
code snippets, along with global state are explained below.


\nop{
\paragraph{Murphi Syntax:} Murphi language consists of {\em rulesets} 
which is a collection of {\em rules}, where each rule is a guarded
action. For example, $SendReqS$ is a ruleset which represents a set of
rules $1..N$, where rule $i$ (denoted by $SendReqS[i]$) corresponds to
$i^{th}$ cache sending a request for shared access. The guard of
$SendReqS[i]$ is $Chan1[i].Cmd = Empty \land Cache[i].State = S$ and
the action is $Chan1[i].Cmd := ReqS$.  {\em Observe that each message
  in the flows shown above correspond to a ruleset in the code.}

\paragraph{Murphi Semantics:} Each rule is executed atomically. Thus, at any time, 
Murphi non-deterministically selects a rule which has its guard true
and executes it.
}

\paragraph{Relevant Global State:} The directory points to the cache
whose request is currently being served (for a particular cache line)
using the pointer \curptr. Further, it stores the request type from
\curptr in \curcmd. Next, \exgntd is set to $true$ if some cache has
\Exclusive access, else it is set to false. Finally, a set of
variables, \shrset, stores the caches which share a particular cache
line. Thus, \shrset[i] is true if cache $i$ has a copy of the
line. Finally, an auxiliary variable $cache\_inv$ (FIX: not shown in
code) points to some cache $c$ such that \shrset[c] is
true---$cache\_inv$ is null if and only if \shrset is false for all
caches.

\paragraph{Code for Exclusive access}

\paragraph{Code for Invalidate access}


\nop{
The flows for the German protocol are shown in
\Fig{figGermanFlows}. These flows represent the order in which various
messages in the protocol are passed, in order to accomplish a high
level request. Thus, they provide high level information about the
protocol. The German protocol consists of the following flows:

\begin{itemize}
\item \Shared: The \Shared flow, shown in \Fig{figGermanFlowsShrd}
  processes a request from cache $i$ for shared access. It consists of
  messages \{$sendReqS$, $RecvReqS$, $SendGntS$, $RecvGntS$\} passed
  between the requesting cache and directory.

\item \Exclusive: Similar to shared flow, shown in
  \Fig{figGermanFlowsExcl}.

\item \Invalidate: The \Invalidate flow is a sub-flow of \Shared and
  \Exclusive flows: it has rules \{$sendInv$, $sendInvAck$,
  $recvInvAck$\}. It is present in both figures,
  \Fig{figGermanFlowsShrd} and \Fig{figGermanFlowsExcl}.
\end{itemize}
}
}

\section{Deriving Invariants for Proving \MDeadlock Freedom}
\label{sec:SpecMethod}
In this section, we show how a set of invariants \InvSet can be
derived from flows such that the invariants in \InvSet collectively
imply \mdeadlock freedom. At a high-level, our method tries to show
\mdeadlock freedom by partitioning the global state of the protocol
using predicates, such that for each partition, some agent $i$ has at
least one transition enabled. Each invariant $inv$ is of the form
$inv.pred \Rightarrow \big(\forall i \in \ISet{inv}: \,
\g(\AgentRuleSet(i))\big)$, where $inv.pred$ is a predicate on the
global variables of the protocol, $\ISet{inv} \subseteq \IndexSet$
s.t. $\lnot(\ISet{inv}=\{\})$ (this is discharged as a separate
assertion for model checking) and $\g(\AgentRuleSet(i))$ denotes a
disjunction of the guards of the rules in $\AgentRuleSet(i)$. {\em The
  key insight is that since $\g(\AgentRuleSet(i))$ has transitions
  from a single agent, the abstractions required for model checking
  $inv$ for an unbounded number of agents are significantly simpler
  than those for checking the original \mdeadlock
  property,\footnote{\revised{This Footnote is new! }In the case of
    rules involving more than one agent (say $c$), the corresponding
    invariants may involve transitions from $c$ agents as well. Since
    $c$ is small for practical protocols, the abstraction constructed
    for verifying such invariants will be simple as well.} as
  discussed in \Sec{sec:CMP}.}

\revised{Mostly Redone!} Our method iteratively model checks each
invariant in \InvSet to refine it. Suppose, the invariant $inv \in
\InvSet$ fails on model checking with the state of the protocol at
failure being $s_f$. Then, there exists some agent $i_f$ such that
when $inv.pred$ holds in $s_f$, $i_f \in \ISet{inv}$ is \true and
$\g(\AgentRuleSet(i_f))$ is \false in $s_f$. This can happen due to
two reasons: first, there may be a mismatch between the flow
specification and the rule-based protocol description. This can be due
to a missing rule in some flow, a missing flow all together, or an
implementation error: the cause for the mismatch can be discovered
from the counterexample. As an example for this case, the
counterexample may show that all flows of the agent $i_f$ are not
enabled, however the agent still has some rule $rl_e(i_f)$ enabled:
this rule may be a part of a missing flow. However, typically the
invariant $inv$ fails due to the second reason: there must exist some
flow \flow of the agent $i_f$ which is blocked (i.e. it has a rule
which is expected to be enabled and so has precondition \true but has
its guard \false). This blocked flow is waiting for another flow
\flow' of another agent $i_s$ to complete. As an example, for the
German protocol, the \Exclusive flow may be blocked for agent $i_f$
with the rule $SendGntE(i_f)$ having precondition \true but guard
\false and waiting for an \Invalidate request to complete for another
agent $i_s$ in the set $Sharers$. In this case, the set \InvSet is
refined by splitting the invariant $inv$.

The invariant $inv$ is split by, (1) splitting the predicate
$inv.pred$ to further partition the global state, and (2) updating the
set $\ISet{inv}$ for each partition. To accomplish this, the user
identifies a pointer variable from $G_P$ or $L_P[i]$ (or an auxiliary
variable) $\ptr$, such that it has the value $i_s$ in the failing
state $s_f$ (and so acts as a {\em witness} variable for $i_s$). The
user also identifies a conflict condition $conf$ on the global state
which indicates when $i_s$ is enabled and $i_f$ fails. This is done by
using the heuristic that if the rule $rl_f(i_f)$ of flow \flow of
agent $i_f$ is blocked, $conf$ can be derived by inspecting the guard
of $rl_f(i_f)$; the condition $conf$ generally is the cause for
falsification of $rl_f(i_f).\rho$.
For example, for the German protocol, $conf$ is derived from the guard
of $SendGntE$ and $\ptr$ points to some sharer which is being
invalidated.

Using $conf$ and $\ptr$, the invariant can be split into two
invariants. (1) The first invariant excludes the case when conflict
happens from the original invariant, i.e., $inv1: (inv.pred \land
\lnot conf) \Rightarrow \big(\forall i \in \ISet{inv1}: \,
\g(\AgentRuleSet(i))\big)$, where $\ISet{inv1} = \ISet{inv}$. (2) The
second invariant shows that when a conflict happens, the agent pointed
to by \ptr must be enabled and so the protocol is still
\mdeadlock-free, i.e., $inv2:$ $(inv.pred \land conf) \Rightarrow
\big(\forall i \in \ISet{inv2} :\, \g(\AgentRuleSet(i))\big)$, where
$\ISet{inv2} = \{i| \, (i \in \IndexSet) \land (i=\ptr)\}$. For both
the invariants, assertions which check that the corresponding set of
indices are non-empty are also verified. For example, for $inv1$, this
assertion is $(inv.pred \land \lnot conf) \Rightarrow \ISet{inv1}$.

\nop{
some rule $rl_e(i_f)$ of the agent $i_f$ is still enabled, even when
the invariant fails. This occurs due to a mismatch between the flow
specification and the protocol implementation---the rule $rl_e(i_f)$
is not present in any flow. This may be either due to an entire flow
missing from the description, or an error in the description of some
flow which missed out $rl_e(i_f)$. Thus, the protocol or the flows
have to be fixed. In the second case,
}

Our method derives these invariants by iteratively model checking with
a small number $c$ ($3$ for German protocol) of agents. (Once the
invariants are derived for $c$ agents, they are verified for an
unbounded number of agents, as shown is \Sec{sec:CMP}.) This number
$c$ needs to be chosen to be large enough such that the proof of
\mdeadlock freedom is expected to generalize to an unbounded number of
agents. For the protocols we verified, we found that as a heuristic,
$c$ should be one more than the maximum number of agents involved in
processing a high-level request. For the German protocol, an
\Exclusive request may involve two agents, a requesting agent $i$ and
an agent $j$ getting invalidated, so we chose $c$ to be equal to 3.

\Fig{fig:MethodOverview} shows the details of the method. It starts
with an initial broad guess invariant, $\true \Rightarrow \big(\forall
i \in \IndexSet: \, \g(\AgentRuleSet(i))\big)$ (line 1). This
indicates that in all reachable states, every agent has at least one
transition enabled. As this invariant is \false, this broad guess
invariant is refined into finer invariants, using the loop. On
finishing, the user is able to derive a set of invariants, $\InvSet$,
which collectively imply \mdeadlock freedom. Further, the user is also
able to derive an assertion set, $\AssertSet$, such that for each
invariant $inv$ in $\InvSet$, an assertion in $\AssertSet$ checks if
the set of indices $\ISet{inv}$ is non-empty when $inv.pred$ holds.

\begin{figure}
\begin{minipage}[b]{\textwidth}
\revised{This Algo is revised}
 {\sc Derive\_Invariants($\Protocol(c)$)}:
      \begin{algorithmic}[1]
        \IndState[0]{$\InvSet = \{ \true \Rightarrow \big(\forall i \in \IndexSet: \, \g(\AgentRuleSet(i))\big)\}$}
        \IndState[0]{$\AssertSet = \{\}$}
        \IndState[0]{while $\Protocol(c) \not\models \InvSet$ do}
        \IndState[0.5]{Let $inv \in \InvSet: \Protocol(c) \not\models inv$ and

                      $inv:$ $inv.pred \Rightarrow \big(\forall i \in \ISet{inv}: \g(\AgentRuleSet(i))\big)$, where, $\ISet{inv} \subseteq \IndexSet$} 

        \IndState[0.5]{Inspect counterexample $cex$ and failing state $s_f$:}
        \IndState[1]{Case 1: mismatch between flows and protocol}
        \IndState[1.5]{Exit loop and fix flows or protocol}
        \IndState[1]{Case 2: identify conflicting agents $i_f$ and $i_s$ s.t.}
        \IndState[1.5]{(1) $i_f: \big((i_f \in \ISet{inv}) \land (\lnot \g(\AgentRuleSet(i_f)))\big)$ holds in $s_f$.}
        \IndState[1.5]{(2) $\exists rl_f \in \flow(i_f)$ s.t. $\big(\precon{\flow}{rl_f}{i_f} \land \lnot(\g(\flow(i_f)))\big)$ holds in $s_f$.}
        \IndState[1.5]{(3) $\g(\AgentRuleSet(i_s))$ holds in $s_f$.}
        \IndState[1]{Identify $conf$ and witness $\ptr$ from above information}
        \IndState[1]{$inv1:$ $(\lnot conf \land inv.pred) \Rightarrow \big(\forall i \in \ISet{inv}: \, \g(\AgentRuleSet(i))\big)$} 
        \IndState[1]{$inv2:$ $(conf \land inv.pred) \Rightarrow \big(\forall i \in \ISet{inv2}: \, \g(\AgentRuleSet(i))\big)$, where,

                       $\ISet{inv2} = \{i| \, i = \ptr\}$}
        \IndState[1]{$\InvSet = \{\InvSet \setminus inv\} \cup \{inv1,inv2\}$}
        \IndState[1]{$\AssertSet = \big(\AssertSet \setminus \big(inv.pred \Rightarrow (\ISet{inv} \neq \{\})\big)\big) \; \cup$ 

                       $\{\big(inv1.pred \Rightarrow (\ISet{inv1} \neq \{\})\big),\big(inv2.pred \Rightarrow (\ISet{inv2} \neq \{\})\big)\}$}
      \end{algorithmic}
\end{minipage} \hspace{-5em} 
\caption{Method for Deriving Invariants from Flows.}
\label{fig:MethodOverview}
\end{figure}

{\bf \emph{Soundness of the Method}} \TechnicalVersion{The following
  theorem shows that the invariants in \InvSet along with the
  assertions in \AssertSet collectively imply \mdeadlock freedom, with
  proof in Appendix~\ref{sec:proof}.}  \CameraReady{The following
  theorem (proof in the extended version~\cite{DJDeadlockPaperArXiv})
  shows that the invariants in \InvSet along with the assertions in
  \AssertSet collectively imply \mdeadlock freedom.}

\nop{\newtheorem*{lem1}{Lemma}
\begin{lem1}
The disjunction of predicates of all invariants in \InvSet holds,
i.e., $\bigvee_{inv \in \InvSet} inv.pred$ holds.
\end{lem1}

This condition intuitively implies that the preconditions of all the
invariants in $\InvSet$ partition the global state of the protocol.
This enables the proof of the following Theorem (proof in
Appendix~\ref{sec:proof}):
}

\newtheorem*{lem3}{Theorem}
\begin{lem3}
If the set of invariants $\InvSet$ along with the set of assertions
\AssertSet hold, they collectively imply \mdeadlock freedom, i.e.,
$\big(\big(\bigwedge_{inv \in \InvSet} (\Protocol \models inv)\big)$
$\land$ $\big(\bigwedge_{asrt \in \AssertSet} (\Protocol \models
asrt)\big)\big)$ $\Rightarrow \big(\Protocol \models (\bigvee_{i}
\bigvee_{j} rl_j(i).\rho)\big)$.
\end{lem3}


\subsection{Specifying Invariants for the German Protocol}
\label{sec:GermanInvariantGeneration}
\revised{This subsection was rewritten} We derive the invariants for a
model of the German protocol with 3 cache agents. We start with the
initial invariant that for all agents, some flow is enabled, i.e.,
\Inv{inv-1}: \(true \Rightarrow \big(\forall i \in \IndexSet: \,
\g(\AgentRuleSet(i))\big)\).

{\bf \emph{Iteration 1:}} Model checking the invariant \Inv{inv-1}
returns a counterexample trace ($SendReqE(1)$, $RecvReqE(1)$,
$SendReqE(2)$). Since the index of the last rule in the trace is $2$,
\g(\AgentRuleSet(2)) must be \false. This is because the rule
$RecvReqE(2)$ of the \Exclusive flow of cache 2 is not fired and thus
has precondition \true but guard \false. The user identifies the
conflict condition $conf = \lnot(CurCmd=Empty)$ from the guard of the
blocked rule $RecvReqE(2)$. Since $CurPtr$ is the witness pointer in
the protocol for the variable $CurCmd$, the witness $\ptr$ is set to
$CurPtr$. Thus, the invariant is split as follows:

\begin{itemize}
\item \Inv{inv-1.1}: \((CurCmd=Empty) \Rightarrow (\forall i \in
  \IndexSet: \, \g(\AgentRuleSet(i)))\).

\item \Inv{inv-1.2}: \(\lnot(CurCmd=Empty) \Rightarrow (\forall i \in
  \ISet{inv-1.2}: \, \g(\AgentRuleSet(i)))\), where $\ISet{inv-1.2} =
  \{i| \, (i \in \IndexSet)\land(i=CurPtr)\}$. The assertion $\lnot
  (CurCmd=Empty)$ $\Rightarrow$ $\lnot (\ISet{inv-1.2}=\{\})$ is also
  checked.
\end{itemize}

{\bf \emph{Iteration 2:}} Next, on model checking the invariants
\Inv{inv-1.1} and \Inv{inv-1.2}, the invariant \Inv{inv-1.2}
fails. The counterexample trace returned is ($SendReqE(1)$,
$RecvReqE(1)$, $SendGntE(1)$, $SendReqE(2)$, $RecvReqE(2)$,
$SendReqE(2)$). Since the last rule of the counterexample is from
cache 2, \g(\AgentRuleSet(2)) must be \false even when
$CurPtr=2$. Further, there are two flows for two \Exclusive requests
by cache 2 active in the counterexample, the first with $SendReqE(2)$
fired and the second with $SendReqE(2)$, $RecvReqE(2)$ fired. Since
the first flow is blocked on the rule $RecvReqE(2)$, the guard of this
rule is inspected. The guard is \false as $CurCmd$ is not
empty. However, since the corresponding witness variable for $CurCmd$
is $CurPtr$ which is already $2$ (due to the processing of the second
flow), this is not a conflict with another cache. The conflict must
then be for the second \Exclusive flow. The second flow is blocked on
the rule $SendGntE(2)$ with precondition \true but guard \false: the
user identifies the conflict condition $conf$ from the guard of
$SendGntE$ to be $Inv\_Cond$. Now, if $Inv\_Cond$ is \true, the
\Invalidate flow for some sharer cache (cache $1$ in this trace) must
be active. Thus, the user identifies $\ptr$ to point to a sharer which
must be invalidated: this is done using the auxiliary variable
$Sharer$, which points to the last sharer to be invalidated in
$ShrSet$. Thus, the invariant \Inv{inv-1.2} is split as follows:

\nop{is due to a conflict with the \Invalidate flow of agent
2: the guard of the rule $SendGntE$ of the \Exclusive flow requires
all sharers to be invalidated, which is what the enabled \Invalidate
of agent 2 is doing . Thus, the invariant inv-1.2 is split by
selecting the witness $\ptr$ to be an auxiliary variable $Sharer$
which points to the last sharer added to the set of sharers, $ShrSet$.
The conflict condition checks if some \Invalidate needs to be sent and
is thus derived from the guard of the $SendInvalidate$ rule and is
equal to $(Inv\_Cond$ $\land$ $\lnot (Sharer=null))$ [CHECK ME]. The
two invariants thus obtained are:
}

\begin{itemize}
\item \Inv{inv-1.2.1}: \(\big(\lnot(CurCmd=Empty) \land (\lnot
  Inv\_Cond)\big) \Rightarrow (\forall i \in \ISet{inv-1.2.1}: \,
  \g(\AgentRuleSet(i)))\), where, \ISet{inv-1.2.1} =
  \ISet{inv-1.2}. An assertion that the precondition implies the index
  set is non-empty is also checked.

\item \Inv{inv-1.2.2}: \(\big(\lnot(CurCmd=Empty) \land
  (Inv\_Cond)\big) \Rightarrow (\forall i \in \ISet{inv-1.2.2}: \,
  \g(\AgentRuleSet(i)))\), where, $\ISet{inv-1.2.2} = \{i| \, (i \in
  \IndexSet)\land(i \in ShrSet)\}$. An assertion that the precondition
  implies the index set is non-empty is also checked.
\end{itemize}

{\bf \emph{Iteration 3:}} Next, on model checking, the invariants
\Inv{inv-1.1}, \Inv{inv-1.2.1}, \Inv{inv-1.2.2}, along with the added
assertions hold for a model with 3 caches. Then, to prove \mdeadlock
freedom, this set of invariants form a candidate set to verify a
protocol model with an unbounded number of agents. The property is
checked for unbounded agents using techniques described in
\Sec{sec:CMP}.

\section{Verifying Flow Properties for Unbounded Agents}
\label{sec:CMP}
We now show how to verify the invariants in $\InvSet$ for an unbounded
number of agents by leveraging the data-type reduction abstraction
along with the CMP method.







{\bf Abstraction: Data-type Reduction}
Since the invariant is of the form $inv.pred \Rightarrow \big(\forall
i \in \ISet{inv}: \g(\AgentRuleSet(i))\big)$, by symmetry, it is
sufficient to check: $inv.pred \Rightarrow \big((1 \in \ISet{inv})
\Rightarrow \big(\g(\AgentRuleSet(1))\big)\big)$. In order to verify
this invariant, just the variables of agent $1$ are required. Then,
our abstraction keeps just the agent $1$, and discards the variables
of all the other agents by replacing them with a state-less {\em
  environment} agent. We refer to agent $1$ as a {\em concrete} agent
and the environment as \Other with {\em id} $o$.

In the original protocol, since all the agents other than agent $1$
interact with it by updating the global variables, the actions of
these agents on the global variables are over-approximated by the
environment agent. This environment agent does not have any local
state. The construction of this agent \Other is automatic and
accomplished syntactically: further details on the automatic
construction are available in \cite{ParamVerMurali}. The final
constructed abstraction then consists of: (1) a concrete agent $1$,
(2) an environment agent \Other with {\em id} $o$, and (3) invariants
specified on variables of agent $1$ and global variables. This
abstraction is referred to as {\em data-type reduction}. If the
original protocol is \Protocol, and invariant set $\InvSet$, we denote
this abstraction by $data\_type$ and thus the abstract model by
$data\_type(\Protocol)$ and the abstracted invariants on agent 1 by
$data\_type(\InvSet)$.

\nop{
In order to prove the invariants for an unbounded number of agents in
the protocol model, a finite model is constructed by using data-type
reduction. This abstraction works by essentially keeping a small
number of agents, as is, and discarding the state of all other
agents. The agents which are left as is are referred to as {\em
  concrete} agents. Leveraging symmetry, suppose agent $1$ is kept as
is and is the concrete agent. Now, 

The agents which are left as is are referred to as {\em
  concrete} agents. Since all the other agents, which are discarded,
interact with the concrete agent by modifying the global state, their
actions on the global state are over-approximated and replaced with an
abstract environment agent, referred to as \Other. The actions of the
environment agent \Other create {\em interference} for the concrete
agents: if the invariants hold despite the interference, then they
hold for the unbounded model as well.  The construction of this
environment agent \Other is automatic and accomplished
syntactically~\cite{ParamVerMurali}.

Next, once the abstraction (with agents $1$, $2$ and \Other) is
constructed, the invariants are proven for the concrete agent $1$ (and if true,
they hold for all agents by symmetry)
}

{\bf \emph{Abstraction for German Protocol}} We now describe how the
rule $SendGntE(i)$ gets abstracted in $data\_type(\Protocol)$. In the
abstract model, there is one concrete agent $1$, which has the rule
$SendGntE(1)$. Next, $SendGntE(o)$ is constructed as follows. (1) The
guard is abstracted by replacing all atoms consisting of local
variables (e.g. $GntChannel[i] = Empty$) with \true or \false
depending on which results in an over-abstraction and by replacing any
usage of $i$ in atoms with global variables (e.g. $CurPtr=i$) with $o$
(i.e. $CurPtr=o$). (2) The action is abstracted by discarding any
assignments to local variables. Further, assignments to global pointer
variables are abstracted as well: any usage of $i$ (e.g. $CurPtr:=i$)
is replaced by $o$ (i.e. $CurPtr:=o$). The rule for agent \Other is
shown below:

\vspace*{0.1cm} 
\begin{minipage}{\textwidth}
\footnotesize
\RulesetSingle{SendGntE($o$)}{CurCmd = ReqE $\land$
  CurPtr = o $\land$ true $\land$ Exgntd = false $\land$ 

\hspace*{.6cm} ShrSet = \{\}}{ShrSet := \{o\}; ExGntd := true; CurCmd :=
  Empty; 

\hspace*{.6cm}CurPtr := NULL;}
\end{minipage}

\nop{
FIX THIS: Next, the invariants to be proven on this abstract model are
modified as well: by symmetry, the invariants only need to be checked
for concrete values of $i$, i.e., $i=\{1,2\}$. Thus, \InvSet is
replaced by \InvConcSet in the above invariants to obtain the
following invariants:

\begin{itemize}
\item \((CurPtr=null) \Rightarrow \g(1)\)

\item \(\lnot(CurPtr=null) \land (1=CurPtr) \land (\lnot Inv\_Cond
  \lor Sharer=null) \Rightarrow \g(1)\)

\item \(\lnot(curptr=null) \land (1 = Sharer) \land (Inv\_Cond \land \lnot
  (Sharer=null)) \Rightarrow \g(1)\)
\end{itemize}
}


\nop{
\newtheorem*{lem4}{Theorem}
\begin{lem4}
Assuming the invariants in $\InvSet$ are single-index and the protocol
$\Protocol$ is symmetric, the abstraction $data\_type$ is sound.
\end{lem4}
We refer an interested reader to \cite{ParamVerMurali} for proof.
}
{\bf The Abstraction-Refinement Loop of the CMP Method}
The CMP method works as an abstraction-refinement loop, as shown in
\Fig{Fig:CMPMethod}. In the loop, the protocol and invariants are
abstracted using data-type reduction. If the proof does not succeed,
the user inspects the returned counterexample $cex$ and following
possibilities arise. (1) Counterexample $cex$ is real, in which case
an error is found and so the loop exits. (2) Counterexample $cex$ is
spurious and so the user refines the protocol by adding a {\em
  non-interference lemma} $lem$. The function {\em strengthen} updates
the guard $rl_j(i).\rho$ of every rule $rl_j(i)$ of the protocol to
$rl_j(i).\rho \land lem(j)$; this way, on re-abstraction with
$data\_type$ in line 1, the new abstract protocol model is
refined. Additional details on the CMP method are available in
\cite{CMP, ParamVerifCMPKristic}.

\nop{
A useful heuristic for determining if the counterexample is spurious
is to examine the behavior of the \Other agent in the counterexample
trace.  Typically, in a spurious trace, the rules of the \Other agent
fire when they are not supposed to (for example doing a memory write
when another agent, say $1$, has exclusive access).
}

\begin{figure}
\centering
    \begin{minipage}{.65\textwidth}
      {\sc CMP}$(\Protocol(N),\InvSet)$
      \begin{algorithmic}[1]        
         \IndState[0]{$\Protocol^{\#}=\Protocol(N)$;$\InvSet^{\#}=\InvSet$}
         \IndState[0]{$while \; data\_type(\Protocol^{\#}) \not\models$ 
$data\_type(\InvSet^{\#}) \; do$}
         \IndState[1]{examine counterexample $cex$}
         \IndState[1]{if $cex$ is real, exit}
         \IndState[1]{if spurious:}
         \IndState[2]{find lemma $lem=\forall i.lem(i)$}
         \IndState[2]{$\Protocol^{\#}=strengthen(\Protocol^{\#},lem)$}
         \IndState[2]{$\InvSet^{\#} = \InvSet^{\#} \cup lem$}
      \end{algorithmic}
    \end{minipage}
\caption{The CMP method}
\label{Fig:CMPMethod}
\end{figure}

\nop{
\newtheorem*{lem5}{Theorem}
\begin{lem5}
Assuming the invariants in $\InvSet$ are single-index and the protocol
$\Protocol$ is symmetric, the CMP loop is sound, i.e.,
$data\_type(\Protocol^{\#}) \models$ $data\_type(\InvSet^{\#})$
$\Rightarrow$ $\Protocol(N) \models$ $\InvSet$ for all values of $N$.
\end{lem5}

We refer an interested reader to \cite{ParamVerMurali} for proof.
}

\nop{
Since the constructed environment agent \Other is completely
unconstrained, on model checking, the property under check may get
violated by the constructed abstract model due to spurious behaviors
exhibited by the agent \Other. The CMP method works by repeatedly
abstracting refined versions of the protocol model until either the
property is proven to be correct, or a real bug is found.

\begin{figure}
  \includegraphics[scale=0.5]{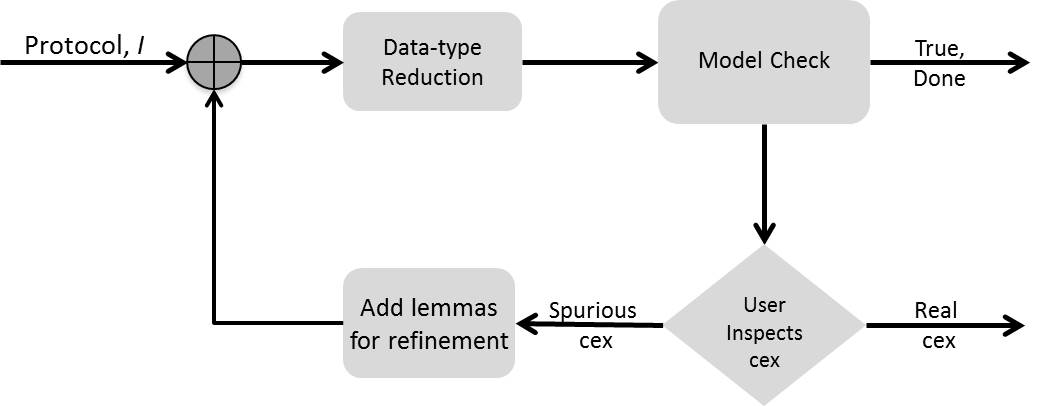}
\caption{The CMP Method}
\label{figCMPLoop}
\vspace{-0.75cm}
\end{figure}

\Fig{figCMPLoop} shows the flow of the CMP method. In the method, the
protocol is first abstracted. Next, model checking of the protocol is
done---in case the model checking succeeds, the property is proven to
be correct for an arbitrary number of agents. If on the other hand the
model checking fails, the user inspects the counter-example.  In case,
the counterexample is a valid counterexample, a real bug has been
found and so the loop exits.  On the other hand, if the counterexample
is spurious, refinement must be done. This is done in two
steps. First, the user comes up with a lemma $lem$ for strengthening
the protocol model. The strengthening is done by adding the lemmas to
the guards of all the rules in the protocol---a rule $rl$ defined as
\(\rho \Rightarrow a\) is strengthened to \(\rho \wedge lem
\Rightarrow a\). Second, re-abstraction is done for the strengthened
protocol: the abstracted version of the rule $rl$ for the thread
\Other is more constrained than the original abstracted version---thus
resulting in a refined abstraction.  Note that during this refinement,
no additional state is added to the model.
This helps in keeping the model tractable for model checking. {\em
  This efficient refinement along with the efficiency of data type
  reduction is the key factor for the success of this method for
  industrial scale protocols.}


}
\nop{as follows: suppose that the lemma $lem$ is used to refine the
  model. Now consider a rule $r$ of the protocol \Protocol defined as:
  \(\rho \Rightarrow a\). Then, refining \Protocol with $lem$ involves
  changing this rule to \(\rho \wedge lem \Rightarrow a\) (we refer to
  this as $strengthen$) and then re-abstracting the new program with
  the new rule obtained.

This abstraction-refinement continues in a loop, as shown in
\Fig{Fig:CMPMethod}. In the loop, whenever the model is verified (line
2), if the proof succeeds, the property is proven (line 9).  If, on
the other hand, there is a counterexample for the refined system, the
user must distinguish between the following cases by examining the
counterexample.  1) The counterexample is valid and so a real bug is
found. (line 4) 2) The counterexample is spurious and so refinement must
be done. (lines 6-8)

A useful heuristic for determining if the counterexample is spurious
is to examine the behavior of the \Other agent in the counterexample
trace.  Typically, in a spurious trace, the rules of the \Other agent
fire when they are not supposed to (for example doing a memory write
when another agent, say 1, has exclusive access).

\begin{figure}
\centering
    \begin{minipage}{.65\textwidth}
      {\sc CMP}$(\Protocol(N),\InvSet)$
      \begin{algorithmic}[1]        
         \IndState[0]{$\Protocol^{\#}=\Protocol(N)$;$\InvSet^{\#}=\InvSet$}
         \IndState[0]{$while \; data\_type(\Protocol^{\#}) \not\models$ 
$data\_type(\InvSet^{\#}) \; do$}
         \IndState[1]{examine counterexample $cex$}
         \IndState[1]{if $cex$ is real, exit}
         \IndState[1]{if spurious:}
         \IndState[2]{find lemma $lem=\forall i.lem(i)$}
         \IndState[2]{$\Protocol^{\#}=strengthen(\Protocol^{\#},lem)$}
         \IndState[2]{$\InvSet^{\#} = \InvSet^{\#} \cup lem$}
      \end{algorithmic}
    \end{minipage}
\caption{The CMP method}
\label{Fig:CMPMethod}
\end{figure}

\paragraph{Note:} Non-interference lemmas were not required for verification
of the invariants we generated for the German protocol; they were only
required for the Flash protocol. Then, for the sake of brevity we skip
the explanation of refinement through an example and refer interested
reader to BLA. [COMPLETE ME]
}

\nop{
\subsection{Advantages of the CMP Method}
The key advantage of the CMP method is that no extra state is added
during refinement: only the guards of rules are strengthened.
This helps in keeping the model tractable for model checkers---this
along with the efficiency of data type reduction is the key factor for
the success of this approach for industrial scale protocols.

Another key advantage of the CMP method is that the lemmas used for
strengthening are also checked during the process and any false lemma
that is added will be detected by the model
checker~\cite{ParamVerifCMPKristic}.  This guarantees that the
refinement step does not affect the soundness of the proof and greatly
enhances its convenience for the user.
}

\nop{ Formally, suppose that the lemma $L$ is used. Now consider a
  rule $r$ of the program $P$ defined as: \(\rho \Rightarrow
  a\). Then, refining $P$ with $L$ involves changing this rule to
  \(\rho \wedge L \Rightarrow a\) and then re-abstracting the new
  program with the new rule obtained. Observe that in this refinement
  approach, no extra state gets added to the abstract model. This is
  important for the efficiency of the CMP method based verification.

This abstraction-refinement continues in a loop, as shown in \Fig{}.  
In the loop, whenever a counterexample is returned by the model checker,
the user determines if the counterexample is real or spurious. 

Whenever a violation 
 On detecting a
violation, the user examines the counter-example.

If this refined system passes the model checker, the property is
proven.  If, on the other hand, there is another counterexample for
the refined system, the user must distinguish between three possible
cases by examining the counterexample.  1) The counterexample is
valid.  2) The counterexample is not valid and the lemma is correct,
in which case further refinement is required.
3) The counterexample is not valid and a lemma is incorrect, in which
case the incorrect lemma must be removed or modified.

}

\section{Experiments}
\label{sec:Experiments}
Using our approach, we verified Murphi (CMurphi 5.4.6) implementations
of the German and Flash protocols (available
online~\cite{ProtocolDeadlockCodeFiles}). Our experiments were done on
a 2.40 GHz Intel Core 2 Quad processor, with 3.74 GB RAM, running
Ubuntu 9.10.

{\bf German Protocol} We verified the invariants discussed in
\Sec{sec:GermanInvariantGeneration}, in order to prove \mdeadlock
freedom. We chose to use an abstraction with 2 agents and an
environment agent, so that the mutual exclusion property can also be
checked.


The proof finished in 217s with 7M states explored. No
non-interference lemmas were required to refine the model, in order to
verify the invariants presented in
\Sec{sec:GermanInvariantGeneration}.  Since typically protocols are
also verified for properties like data integrity (i.e. the data stored
in the cache is consistent with what the processors intended to write)
and mutual exclusion, we model checked the above invariants along with
these properties.  In this case, the abstract model was constrained
and model checking this model was faster and took 0.1 sec with 1763
states explored.

{\bf \emph{Buggy Version}} We injected a simple error in the German
protocol in order to introduce an \mdeadlock. In the bug, an agent
being invalidated drops the acknowledgement $SendInvAck$ it is
supposed to send to the directory. This results in the entire protocol
getting blocked, hence an \mdeadlock situation. This was detected by
the failing of the invariant \Inv{inv-1.2.2}, discussed in
\Sec{sec:GermanInvariantGeneration}.

{\bf Flash Protocol} Next, we verified the Flash
protocol~\cite{FlashProtocol} for deadlock freedom. The Flash protocol
implements the same high-level requests as the German protocol. It
also uses a directory which has a Boolean variable $Pending$ which is
\true if the directory is busy processing a request from an agent
pointed to by another variable $CurSrc$ (name changed from original
protocol for ease of presentation). However, the Flash protocol uses
two key optimizations over the German protocol. First, the Flash
protocol enables the cache agents to directly forward data between
each other instead of via the directory, for added speed. This is
accomplished by the directory by forwarding incoming requests from the
agent $i$ to the destination agent, $FwDst(i)$, with the relevant
data. Second, the Flash protocol uses non-blocking invalidates, i.e,
the \Exclusive flow does not have to wait for the \Invalidate flow to
complete for the sharing agents in $ShrSet$. Due to these
optimizations, the flows of the Flash protocol are significantly more
complex than those of German protocol. Further, due to forwarding,
some rules involve two agents instead of one for the German protocol:
thus the flows involve two agents as well. Each flow then is of the
form $\flow_k(i,j)$, where $i$ is the requesting agent for a flow and
$j=FwDst(i)$ is the destination agent to which the request may be
forwarded by the directory. Then, we define $\AgentRuleSet(i)$ to be
equal to $\bigcup_k\flow_k(i,FwDst(i))$.

We derived the invariants from the flows by keeping $c$ to be equal to
3, as each request encompasses a maximum of 2 agents (forwarding and
invalidation do not happen simultaneously in a flow). The final
invariants derived using our method are as follows:

{\bf \emph{Directory Not Busy:}} If the directory is not busy (i.e.,
$Pending$ is \false), any agent $i$ can send a request. Thus the
invariant \Inv{invF-1}: \(\lnot(Pending) \Rightarrow \big(\forall i
\in \IndexSet: \g(\AgentRuleSet(i))\big).\)

However, if the directory is busy (i.e., $Pending$ is \true), two
possibilities arise. (1) It may be busy since it is processing a
request from agent $CurSrc$. Or, (2) in case the request from $CurSrc$
requires an invalidate, the directory may remain busy with
invalidation even after the request from $CurSrc$ has been
served. This is because Flash allows the request from $CurSrc$ to
complete before invalidation due to non-blocking invalidates. Hence
the following invariants:

{\bf \emph{Directory Busy with Request:}} Invariant \Inv{invF-2}:
\(\big((Pending) \land (ShrSet=\{\})\big) \Rightarrow \big(\forall i
\in \ISet{\Inv{invF-2}} \g(\AgentRuleSet(i))\big),\) where
$\ISet{\Inv{invF-2}} = \{i| \, (i \in \IndexSet) \land (i=CurSrc)\}$.

{\bf \emph{Directory Busy with Invalidate:}} Invariant \Inv{invF-3}:
\(\big((Pending) \land \lnot (ShrSet=\{\})\big) \Rightarrow
\big(\forall i \in \ISet{invF-3} \g(\AgentRuleSet(i))\big),\) where
$\ISet{\Inv{invF-3}} = \{i| \, (i \in \IndexSet) \land (i \in
ShrSet)\}$.

{\bf \emph{Runtime:}} We verified the above invariants
along with the mutual exclusion and the data integrity properties for
an unbounded model abstracted by keeping 3 concrete agents (one agent
behaves as a directory) and constructing an environment agent
\Other. The verification took 5127s with about 20.5M states and 152M
rules fired. In this case we reused the lemmas used in prior work by
Chou \etal~\cite{CMP} for verifying the mutual exclusion and data
integrity properties in order to refine the agent \Other.

{\bf \emph{Verifying Flash vs German Protocol:}} \revised{This
  sub-section} The flows of the Flash protocol involve two indices: we
eliminated the second index by replacing it with the variable
$FwDst(i)$ which stores information of the forwarded cache and thus
made the verification similar to the German protocol case. Next, Flash
protocol uses lazy invalidate: even if the original request has
completed, the directory may still be busy with the invalidate. As
explained above, this was in contrast to the German protocol and
resulted in an additional invariant \Inv{invF-3}.

{\bf \emph{Comparison with Other Techniques:}} \revised{This
  sub-section} The only technique we are aware of which handles Flash
with a high degree of automation is by Bingham
\etal~\cite{DeadlockBingham}. While a direct comparison of the runtime
between their approach and ours is infeasible for this paper, we note
that the invariants generated using our approach only require an
over-abstraction in contrast to theirs which requires a
mixed-abstraction. This is an advantage since development of automatic
and scalable over-abstraction based parameterized safety verification
techniques is a promising area of ongoing research
(e.g. \cite{ProtocolFlashAmitGoel}) which our approach directly
benefits from.

\nop{ {\bf \emph{Verification Experience:}} The Flash protocol
  presented us with unique challenges during verification. First, the
  usage of forwarding made reasoning about Flash hard: as discussed
  above, $\AgentRuleSet(i)$ is single index (denotes requests sent by
  agent $i$). However, since, some of the transitions in
  $\AgentRuleSet(i)$ are two index (second index corresponds to
  $FwDst(i)$), the above invariants really are two index invariants,
  i.e., of the form $\forall i,j: \phi(i,j)$. We had to take this into
  account for verifying for unbounded number of agents.

Next, as discussed above, due to the non-blocking invalidate, when the
failure happens due to an invalidate in progress, the conflict
condition $conf$ is still $Pending$; we used $sharer=null$ as $conf$,
where $sharer$ is an auxiliary variable. This is because, unlike
German protocol, Flash sends the grant without waiting for invalidate
to complete. However, the directory keeps $Pending$ \true and waits
for invalidate to complete, before serving fresh requests, thus
blocking multiple flows and making the identification of $conf$
difficult.
}

\nop{

 The request from the first agent, may get forwarded to the
second agent. We note that in this case, the second agent is pointed
to by an auxiliary pointer $CurDst$.

Thus, for showing \mdeadlock freedom, we start with initially: \(\true
\Rightarrow \forall i \in \IndexSet,j \in \IndexSet:
\g(\AgentRuleSet(i,j))\). Since a maximum of two agents are involved in
the flows for the Flash protocol, we kept $c$ to be equal to $3$.

On model checking, invariant inv-1 failed trivially due to the trivial
case that the agent $j$ gets randomly assigned. So we split the invariant indicating
that if CurDst is not null, $j$ must be $CurDst$.

--- \(\lnot (CurDst = null) \Rightarrow \forall i \in \IndexSet,(j = CurDst):
\g(\AgentRuleSet(i,j))\)

and 

\((CurDst = null) \Rightarrow \forall i \in \IndexSet, j \in
\IndexSet: \g(\AgentRuleSet(i,j))\)

Then, the invariant failed because of pending. So, new invariant
became \((CurDst = null) \land (pending) \Rightarrow \forall i =CurSrc, j
\in \IndexSet: \g(\AgentRuleSet(CurSrc,j))\)

and

--- became \((CurDst = null) \land (!pending) \Rightarrow \forall i, j
\in \IndexSet: \g(\AgentRuleSet(i,j))\)

Next, failed because agent sharer was getting invalidated. Conflict condition
tricky as invalidate is lazy - so only pending is true. So another check that
if there is a sharer, it must be getting invalidated. So split.

--- \((CurDst = null) \land (pending) \land (sharer=null)\Rightarrow \forall i =CurSrc, j
\in \IndexSet: \g(\AgentRuleSet(CurSrc,j))\)

--- \((CurDst = null) \land (pending) \land \lnot(sharer=null)\Rightarrow \forall i = Sharer, j
\in \IndexSet: \g(\AgentRuleSet(CurSrc,j))\)

}

\nop{
and we iterated on refining
the invariant, as in \Sec{Bla}[FILL]. The counterexamples in
refinement showed: (1) that the directory points to the agent whose
flow is being processed using the pointer $CurSrc$~\footnote{Name
  changed from publicly available protocol for clarity.} and (2) in
case of requests from $CurSrc$ which directly get forwarded to another
agent, the destination agent is stored in $CurDst$. Finally, the
sharer of the cache line is stored in the variable $Sharer$

Thus, the final generated invariants are:

\begin{itemize}
\item $CurSrc=null \rightarrow \forall i,j: \g(i,j)$

\item $\lnot(CurSrc=null) \land \lnot(CurDst = null) \rightarrow \g(CurSrc,CurDst)$

\item $\lnot(CurSrc=null) \land (CurDst) = null) \land (Sharer = null)$  \\
$\rightarrow \forall j: \g(CurSrc,j)$

\item $\lnot(CurSrc=null) \land (CurDst) = null) \land  \lnot(Sharer = null)$ \\ 
 $\rightarrow \forall j: \g(Sharer,j)$
\end{itemize}

}

\nop{
The first invariant states that if the forward destination
  stored in $CurDst$ is not $null$, the flow with source $CurSrc$ and
  destination $CurDst$ must be enabled. This is stated as the
  invariant:
  \[\lnot(CurSrc=null) \land \lnot(CurDst = null) \rightarrow
  \g(CurSrc,CurDst)\tag{inv-1.2.1}.\]

  \[\lnot(CurSrc=null) \land (CurDst) = null) \land
  (Sharer = null) \\ \rightarrow \forall j:
  \g(CurSrc,j)\tag{inv-1.2.2.1}.\]

  \[\lnot(CurSrc=null) \land (CurDst) = null) \land  \lnot(Sharer = null) \\ 
 \rightarrow \forall j: \g(Sharer,j)\tag{inv-1.2.2.2}.\]
}

\nop{
\begin{enumerate}
\item The first invariant states that if $CurSrc$ is $null$, then some
  flow for any source agent $i$ and destination agent $j$ is
  enabled. This was stated as: \[CurSrc=null \rightarrow \forall i,j:
  \g(i,j)\tag{inv-1.1}.\]

\item The second invariant states that if $CurSrc$ is not null, then
  some flow with source $CurSrc$ and any destination $j$ is
  enabled. This was stated as:
  \[\lnot CurSrc=null \rightarrow \forall j: \g(CurSrc,j)\tag{inv-1.2}.\]
\end{enumerate}

On model checking, invariant inv-1.2 failed, as many flows have a
single destination agent and thus for other destination agents, no
flow may be enabled.  Information about the destination node to which
a request is sent (or forwarded) to, is stored in the variable
$CurDst$. Then, we split inv-1.2 into two invariants as follows:

\begin{enumerate}
\item The first invariant states that if the forward destination
  stored in $CurDst$ is not $null$, the flow with source $CurSrc$ and
  destination $CurDst$ must be enabled. This is stated as the
  invariant:
  \[\lnot(CurSrc=null) \land \lnot(CurDst = null) \rightarrow
  \g(CurSrc,CurDst)\tag{inv-1.2.1}.\]

\item The second invariant states that if the forward destination
  value stored in $CurDst$ is $null$, all flows with source $CurSrc$
  with any destination must be enabled. This is specified as the
  invariant:
   \[\lnot(CurSrc=null)\land (CurDst = null) \rightarrow \forall
  j: \g(CurSrc,j)\tag{inv-1.2.2}.\]
\end{enumerate}

On model checking the above invariants, the invariant inv-1.2.2
failed: as in this case, at times an \Invalidate is happening to
another agent and thus the request is not enabled. The last added
sharer for the cache line is pointed to by the witness variable
$Sharer$. Then, we split the invariant inv-1.2.2 into two invariants:

\begin{enumerate}
\item The first invariant states that if the last sharer which needs
  to be invalidated is $null$ then any flow with source $CurSrc$ must
  be enabled. This is stated as the invariant: 
\begin{align*}
  \hspace*{-.4cm}\lnot(CurSrc=null) \land (CurDst) = null) \land
  (Sharer = null) \\ \rightarrow \forall j:
  \g(CurSrc,j)\tag{inv-1.2.2.1}.
\end{align*}

\item The second invariant states that if the last sharer is not
  $null$, then the some flow (i.e. the \Invalidate flow) for the last
  sharer must be enabled. This is stated as the invariant:
\begin{align*}
  \hspace*{-.4cm} \lnot(CurSrc=null) \land (CurDst) = null) \land  \lnot(Sharer = null) \\ 
 \rightarrow \forall j: \g(Sharer,j)\tag{inv-1.2.2.2}.
\end{align*}
\end{enumerate}

On model checking with 3 agents, invariants inv-1.1, inv-1.2.1,
inv-1.2.2.1, inv-1.2.2.2 were verified to be $true$ for the Flash
protocol. Since these agents involve a small number of indices, they
were verified for an unbounded number of indices using the CMP method.
}


\nop{
Next, in order to do a sanity check and see if non-interference lemmas
are required for verifying \Enabled alone, we verified just the
\Enabled property (without mutual exclusion and data integrity
properties). In this case we had to add some non-interference lemmas
constraining the interference from the abstract agent (not detailed
for brevity). However, since properties like data integrity were not
being checked, the agent had to be constrained lesser. Thus, the
verification took 48230 sec to finish with 99M states explored and
1935M rules fired.
}

\nop{

Verification of flash protocol for unbounded agents was done using the
CMP method. [Non-interference lemmas?]

----non-interference lemmas same as with mutex.

\subsubsection{Runtime}
The running time for the verification of the flash protocol was 

-- with mutual exclusion
20556798 states, 152413697 rules fired in 4830.49s.

-- without mutual exclusion
98844472 states, 1935368444 rules fired in 48230.24s.
}

\nop{
\section{Discussion: Livelock verification for German Protocol}
We now discuss how our approach can be extended for proving livelock
freedom for the German protocol example. The flows for the German
protocol have characteristics which simplify the definition and
verification of livelock--we leave a general definition and
verification of livelock freedom to future work.

The flows for German protocol have following key properties:

\begin{enumerate}

\item The flows for German protocol are acyclic. This property is also
  true for flows of the protocols we have seen so far.

\item All flows for the German protocol, on completion, map to a
  request successfully served. This is not $true$ for other protocols
  in general: for flash for example, flows may finish due to a
  transition representing failure to grant permission and so a
  high-level request is not served.

\end{enumerate}

Due to the above key properties, it is straightforward to see that if,
at all times, progress can happen along a flow, i.e. some guard of
some flow is always enabled and on firing, the next message in the
flow is the one which can become enabled, then, livelock can not
exist.

First, we strengthen the definition of a flow. For simplicity, we
define a flow to be a total order of messages (in general it is a
partial order however this is enough for German flow).

Thus, $fl$: $\{rl_1,rl_2,\ldots,rl_n\}$, where $rl_{i+1}$ must happen
after $rl_i$. Further, we say that a flow $fl$ is in state $i$ if the
rule $fl_i$ is the only rule which can be enabled. Thus the flow
starts from state 1.

Further, for each agent, a maximum of $n$ instances of flow $fl$ can
exist, due to bounded state.

Then, let $aux_{fl}$ track the state of a flow and let it be
updated in every rule $rl_i$:

1) progress for a flow:

$progress(fl(i))$ = $\bigwedge_i$ $(aux_{fl}=i)$ $\rightarrow$ $rl_i.g$

Then, progress of an agent $i$ is: $progress(i) = \bigvee_{fl(i)}
progress(fl(i))$, where $fl(i)$ is an enabled flow in $i$ (note that
there can be multiple enabled flow of the same type in $i$).

Then, we can show the progress property: i.e. that at all times, there
exists an agent, such that $progress(i)$ holds.

Claim: progress implies livelock freedom.

Intuition: Due to acyclic nature of flows, this is the case.

\subsubsection{Runtime}
The verification of the progress property for the German protocol took
about 97 sec, with 14.5M states explored and 9.2M rules fired.

}

\section{Conclusions and Future Work}
In this paper we have presented a method to prove freedom from a
practically motivated deadlock error which spans the entire cache
coherence protocol, an \mdeadlock.  Our method exploits high-level
information in the form of message sequence diagrams---these are
referred to as {\em flows} and are readily available in industrial
documents as charts and tables. Using our method, a set of invariants
can be derived which collectively imply \mdeadlock freedom. These
invariants enable the direct application of industrial scale techniques
for parameterized verification.

As part of future work, we plan to take up verification of livelock
freedom by exploiting flows. Verifying livelock requires formally
defining a notion of the protocol doing useful work. This information
is present in flows---efficiently exploiting this is part of our
ongoing research.

\bibliographystyle{splncs03.bst}
\bibliography{References}
\newpage
\newpage
\appendix
\section{The German Protocol Code (Chou \etal~\cite{CMP})}
\label{AppendixGermaCode}

\begin{minipage}[t]{\textwidth}\tiny
\hspace*{-0.1\textwidth}
\begin{tabular}[t]{ll}
\begin{minipage}[t]{0.4\textheight}
\begin{verbatim}
const  ---- Configuration parameters ----

  NODE_NUM : 4;
  DATA_NUM : 2;

type   ---- Type declarations ----

  NODE : scalarset(NODE_NUM);
  DATA : scalarset(DATA_NUM);

  CACHE_STATE : enum {I, S, E};
  CACHE : record State : CACHE_STATE; Data : DATA; end;

  MSG_CMD : enum {Empty, ReqS, ReqE, Inv, InvAck, GntS, GntE};
  MSG : record Cmd : MSG_CMD; Data : DATA; end;

var   ---- State variables ----

  Cache : array [NODE] of CACHE;      -- Caches
  Chan1 : array [NODE] of MSG;        -- Channels for Req*
  Chan2 : array [NODE] of MSG;        -- Channels for Gnt* and Inv
  Chan3 : array [NODE] of MSG;        -- Channels for InvAck
  InvSet : array [NODE] of boolean;   -- Nodes to be invalidated
  ShrSet : array [NODE] of boolean;   -- Nodes having S or E copies
  ExGntd : boolean;                   -- E copy has been granted
  CurCmd : MSG_CMD;                   -- Current request command
  CurPtr : NODE;                      -- Current request node
  MemData : DATA;                     -- Memory data
  AuxData : DATA;                     -- Latest value of cache line

---- Initial states ----

ruleset d : DATA do startstate "Init"
  for i : NODE do
    Chan1[i].Cmd := Empty; Chan2[i].Cmd := Empty; Chan3[i].Cmd := Empty;
    Cache[i].State := I; InvSet[i] := false; ShrSet[i] := false;
  end;
  ExGntd := false; CurCmd := Empty; MemData := d; AuxData := d;
end end;

---- State transitions ----

ruleset i : NODE do rule "SendReqS"
  Chan1[i].Cmd = Empty & Cache[i].State = I
==>
  Chan1[i].Cmd := ReqS;
end end;

ruleset i : NODE do rule "SendReqE"
  Chan1[i].Cmd = Empty & (Cache[i].State = I | Cache[i].State = S)
==>
  Chan1[i].Cmd := ReqE;
end end;

ruleset i : NODE do rule "RecvReqS"
  CurCmd = Empty & Chan1[i].Cmd = ReqS
==>
  CurCmd := ReqS; CurPtr := i; Chan1[i].Cmd := Empty;
  for j : NODE do InvSet[j] := ShrSet[j] end;
end end;

ruleset i : NODE do rule "RecvReqE"
  CurCmd = Empty & Chan1[i].Cmd = ReqE
==>
  CurCmd := ReqE; CurPtr := i; Chan1[i].Cmd := Empty;
  for j : NODE do InvSet[j] := ShrSet[j] end;
end end;
\end{verbatim}
\end{minipage}
&
\begin{minipage}[t]{0.4\textheight}
\begin{verbatim}
ruleset i : NODE do rule "SendInv"
  Chan2[i].Cmd = Empty & InvSet[i] = true &
  ( CurCmd = ReqE | CurCmd = ReqS & ExGntd = true )
==>
  Chan2[i].Cmd := Inv; InvSet[i] := false;
end end;

ruleset i : NODE do rule "SendInvAck"
  Chan2[i].Cmd = Inv & Chan3[i].Cmd = Empty
==>
  Chan2[i].Cmd := Empty; Chan3[i].Cmd := InvAck;
  if (Cache[i].State = E) then Chan3[i].Data := Cache[i].Data end;
  Cache[i].State := I; undefine Cache[i].Data;
end end;

ruleset i : NODE do rule "RecvInvAck"
  Chan3[i].Cmd = InvAck & CurCmd != Empty
==>
  Chan3[i].Cmd := Empty; ShrSet[i] := false;
  if (ExGntd = true)
  then ExGntd := false; MemData := Chan3[i].Data; undefine Chan3[i].Data end;
end end;

ruleset i : NODE do rule "SendGntS"
  CurCmd = ReqS & CurPtr = i & Chan2[i].Cmd = Empty & ExGntd = false
==>
  Chan2[i].Cmd := GntS; Chan2[i].Data := MemData; ShrSet[i] := true;
  CurCmd := Empty; undefine CurPtr;
end end;

ruleset i : NODE do rule "SendGntE"
  CurCmd = ReqE & CurPtr = i & Chan2[i].Cmd = Empty & ExGntd = false &
  forall j : NODE do ShrSet[j] = false end
==>
  Chan2[i].Cmd := GntE; Chan2[i].Data := MemData; ShrSet[i] := true;
  ExGntd := true; CurCmd := Empty; undefine CurPtr;
end end;

ruleset i : NODE do rule "RecvGntS"
  Chan2[i].Cmd = GntS
==>
  Cache[i].State := S; Cache[i].Data := Chan2[i].Data;
  Chan2[i].Cmd := Empty; undefine Chan2[i].Data;
end end;

ruleset i : NODE do rule "RecvGntE"
  Chan2[i].Cmd = GntE
==>
  Cache[i].State := E; Cache[i].Data := Chan2[i].Data;
  Chan2[i].Cmd := Empty; undefine Chan2[i].Data;
end end;

ruleset i : NODE; d : DATA do rule "Store"
  Cache[i].State = E
==>
  Cache[i].Data := d; AuxData := d;
end end;

---- Invariant properties ----

invariant "CtrlProp"
  forall i : NODE do forall j : NODE do
    i != j -> (Cache[i].State = E -> Cache[j].State = I) &
              (Cache[i].State = S -> Cache[j].State = I | Cache[j].State = S)
  end end;

invariant "DataProp"
  ( ExGntd = false -> MemData = AuxData ) &
  forall i : NODE do Cache[i].State != I -> Cache[i].Data = AuxData end;
\end{verbatim}
\end{minipage}
\end{tabular}
\end{minipage}
\\ ~ \\ ~

\section{Proof of Soundness}
\label{sec:proof}

Before proving the theorem, we first establish the following lemma:

\newtheorem*{lem1}{Lemma}
\begin{lem1}
The disjunction of predicates of all invariants in \InvSet holds,
i.e., $\bigvee_{inv \in \InvSet} inv.pred$ holds.
\end{lem1}

\begin{proof}
We prove this by induction over the splitting step in our method.

{\em Base Case:} Our method starts with the initial invariant
$true \Rightarrow \big(\forall i \in
IndexSet: \g(\AgentRuleSet(i))\big)$ in \InvSet. Thus, it trivially
satisfies the lemma.

\revised{The proof here}
{\em Induction Step:} Next, suppose at some point during the
generation of invariants, the set of candidates is \InvSet. On model
checking, invariant $inv$ in \InvSet fails with
$\g(\AgentRuleSet(i_f))$ being \false for agent $i_f$. In case there
is an error in the protocol or flows due to a rule $rl(i_f)$ being
enabled for agent $i_f$ in the failing state, the loop exits without
modifying \InvSet and so the lemma holds trivially. In the second
case, the invariant is split into invariants $inv1$ and $inv2$ by
using conflict condition $conf$.


Now for this case, $inv1.pred = (inv.pred \land \lnot conf)$ and
$inv2.pred = (inv.pred \land conf)$. Clearly, the disjunction of
predicates $inv1$ and $inv2$ equals to $inv.pred$, the predicate of
$inv$. Thus, the disjunction of predicates of the new and old set of
invariants is the same, i.e., $\bigvee_{inv \in \InvSet} inv.pred$ =
$\bigvee_{inv \in \InvSet'} inv.pred$, where the new set of invariants
$\InvSet'$ $=$ $\InvSet$ $\setminus$ $\{inv\}$ $\cup$ $\{inv1,inv2\}$.

Hence, by induction, the above lemma holds. 
\end{proof}

Now, using the above lemma, we prove the following theorem to
establish soundness of our method:


\begin{lem3}
If the set of invariants $\InvSet$ along with the set of assertions
\AssertSet hold, they collectively imply \mdeadlock freedom, i.e.,
$\big(\big(\bigwedge_{inv \in \InvSet} (\Protocol \models inv)\big)$
$\land$ $\big(\bigwedge_{asrt \in \AssertSet} (\Protocol \models
asrt)\big)\big)$ $\Rightarrow \big(\Protocol \models (\bigvee_{i}
\bigvee_{j} rl_j(i).\rho)\big)$.
\end{lem3}

\begin{proof}
Let the protocol be in some reachable state $s$. We argue that some
agent has at least one rule enabled in every such reachable state. By
the above lemma, $\bigvee_{inv \in \InvSet} inv.pred$ holds in state
$s$. Thus, there must exist some invariant $inv$ such that its
predicate holds in $s$, i.e., $\exists inv \in \InvSet: inv.pred =
true$.

Now, let $inv$ be $inv.pred \Rightarrow (\forall
i \in \ISet{inv}: \, \g(\AgentRuleSet(i)))$. Then, since the assertion
$inv.pred \Rightarrow \ISet{inv} \neq \{\}$ is in the set \AssertSet,
which holds as well, there is some agent $i_0$ such that it is in
$\ISet{inv}$ and $\g(\AgentRuleSet(i_0))$ holds, i.e., $\exists
i_o \in \ISet{inv}: \, \g(\AgentRuleSet(i_0))$.  Thus, agent $i_0$ is
enabled in the state $s$, and so the state is not an \mdeadlock state.


\end{proof}

\end{document}